\documentclass[11pt,english,%
titlepage,%
]{report}
\usepackage[letterpaper,left=1in,right=1in,top=1in,bottom=1in]{geometry}
\usepackage[utf8]{inputenc}

\usepackage{graphicx}
\usepackage{multirow}
\usepackage{import}
\usepackage{standalone}
\usepackage{siunitx}
\usepackage{fancyhdr}
\usepackage{pdfpages}
\usepackage{amsmath,amssymb,amsthm}
\usepackage{titlesec}
\usepackage{enumitem}
\usepackage{wrapfig}
\usepackage{subcaption}
\usepackage{float}
\usepackage{booktabs}
\usepackage{xcolor}
\usepackage{dirtree}
\usepackage{tikz}
\usetikzlibrary{arrows.meta,decorations.pathreplacing,calc}

\definecolor{sunfill}{RGB}{255,236,204}
\definecolor{sunedge}{RGB}{201,124,20}
\definecolor{roi}{RGB}{31,102,168}
\definecolor{ext}{RGB}{176,42,88}
\definecolor{impl}{RGB}{118,118,126}
\definecolor{ink}{RGB}{40,44,52}

\let\origunderscore\_
\renewcommand{\_}{\origunderscore\allowbreak}
\newcommand\arcsec{\mbox{$^{\prime\prime}$}}
\newcommand{\classlabel}[1]{\label{cls:#1}}
\newcommand{\clsentry}[1]{\hyperref[cls:#1]{\texttt{#1}}\ \ (\S\ref{cls:#1}, p.~\pageref{cls:#1})}

\usepackage{biblatex}
\usepackage{url}
\usepackage{hyperref}
\hypersetup{
    colorlinks=true,
    linkcolor=blue,
    filecolor=magenta,
    urlcolor=cyan,
}

\usepackage{authblk}

\title{\LARGE Describing the \texttt{swdatatoolkit}:\\[0.3em]
    A Space Weather Data Analysis Library\par}
\author[1]{Dustin Kempton}
\author[2]{Griffin Goodwin}
\author[1]{Tarun Kumar Reddy Thippareddy}
\author[1]{Reet Gupta}
\author[2]{Viacheslav Sadykov}
\author[1]{Rafal Angryk}
\affil[1]{Department of Computer Science, Georgia State University}
\affil[2]{Department of Physics \& Astronomy, Georgia State University}
\date{\today}

\begin{document}
\maketitle
\vspace{1.5cm}

\pagenumbering{roman}
\setcounter{page}{1}
\tableofcontents
\fancypagestyle{plain}{%
\fancyfoot[C]{Table of Contents}
}
\newpage

\pagenumbering{arabic}
\setcounter{page}{1}

\fancyfoot[C]{Spaceweather Data Toolkit}
\fancyfoot[R]{Page \thepage}

\fancypagestyle{plain}{%
\fancyfoot[C]{Spaceweather Data Toolkit}
\fancyfoot[R]{Page \thepage}
}

%==========================================================
\chapter*{Abstract}
%==========================================================

\href{https://dmlab.cs.gsu.edu/docs/spaceweather_toolkit/}{\textbf{swdatatoolkit}} is a Python-based scientific software library designed to support the acquisition, preprocessing, and analysis of solar and space weather data. The toolkit consolidates functionality across multiple domains, including data downloading from established heliophysics sources, image preprocessing, edge detection, image texture quantification, magnetic field analysis, and the derivation of higher-level parameters commonly used in solar physics research. Its modular structure reflects the heterogeneous nature of space weather data and enables reproducible, extensible workflows for both exploratory analysis and machine-learning-driven studies. This paper presents an overview of the library's available capabilities, its scientific motivations, and its role in the broader space weather research ecosystem.

%==========================================================
\chapter{Introduction}
%==========================================================

Space weather research requires the integration of diverse observational products from solar telescopes, space-borne instruments, and cataloged event repositories. Investigators typically combine image data, magnetograms, flare records, particle measurements, and contextual metadata to study solar activity and to construct predictive models of eruptive behavior. The
technical burden of acquiring, organizing, and transforming such data is substantial. \href{https://dmlab.cs.gsu.edu/docs/spaceweather_toolkit/}{\textbf{swdatatoolkit}} addresses this challenge by providing a unified library of routines for space weather data ingestion and feature generation.\\
\\

\noindent The library is particularly oriented toward active-region analysis and flare-related studies, where the quality and consistency of derived features strongly influence scientific interpretation and model performance. Rather than serving as a single-purpose utility, \href{https://dmlab.cs.gsu.edu/docs/spaceweather_toolkit/}{\textbf{swdatatoolkit}} functions as a general-purpose research infrastructure for solar data science.

%==========================================================
\chapter{Library Overview}
%==========================================================

\href{https://dmlab.cs.gsu.edu/docs/spaceweather_toolkit/}{\textbf{swdatatoolkit}} is organized into several domain-specific subpackages. Each subpackage addresses a distinct stage in the analytical workflow, from raw data retrieval to feature derivation. The major areas of functionality include:

\begin{enumerate}
    \item Data source access and downloading
    \item Solar image preprocessing
    \item Edge detection and image-structure analysis
    \item Image texture and statistical parameter calculation
    \item Magnetic field computation
    \item Magnetic-parameter derivation
    \item Slicing of time series into labelled samples and feature extraction for analytical pipelines
\end{enumerate}
This modular design is appropriate for scientific software because it allows users to adopt only the components relevant to a particular study while preserving interoperability between components.

%////////////////////Section\\\\\\\\\\\\\\\\\\\\\\\\\\\\\\
\section{Package Structure and Class Index}\label{sec:class_index}

The tree below shows the package hierarchy of \href{https://dmlab.cs.gsu.edu/docs/spaceweather_toolkit/}{\textbf{swdatatoolkit}}, laid out like the module tree of the documentation website. Each class is listed with the section and page where it is described, and every entry is a hyperlink to that section. Enumerations and helper classes are listed next to the class that uses them.

% The tree is split in two so that each half fits on a page (dirtree cannot break across pages).
\noindent\begin{minipage}{\textwidth}
\small
\DTsetlength{0.2em}{1em}{0.2em}{0.4pt}{1.6pt}
\dirtree{%
.1 \textbf{swdatatoolkit}.
.2 \textbf{datasources} --- data acquisition (Chapter~\ref{chap:datasources}).
.3 \textit{Image and active-region retrieval}.
.4 \clsentry{AIADownloader}; uses \texttt{WaveBand}.
.4 \clsentry{HARPDownloader}; uses \texttt{HARPDataType}.
.3 \textit{Event and catalog downloaders}.
.4 \clsentry{AIAFlareEventDownloader}.
.4 \clsentry{GOESFlareEventDownloader}.
.4 \clsentry{SRSDownloader}.
.4 \clsentry{SSWEventDownloader}.
.4 \clsentry{CMEDownloader}; uses \texttt{CACTUSReportType}, \texttt{ReportCat}.
.4 \clsentry{XRTHinodeEventDownloader}.
.3 \textit{Time-series downloaders}.
.4 \clsentry{SXRDownloader}.
.4 \clsentry{ProtonDownloader}.
.4 \clsentry{OMNIWebDownloader}.
.4 \clsentry{SOHOEPHINDownloader}.
.4 \clsentry{NeutronDownloader}.
.4 \clsentry{DSCOVRDownloader}.
.2 \textbf{imageproc} --- image preprocessing (Chapter~\ref{chap:imageproc}).
.3 \clsentry{AIACutOutProcessor}.
.2 \textbf{edgedetection} --- edges and gradients (Chapter~\ref{chap:imageproc}).
.3 \clsentry{BaseEdgeDetector}.
.3 \clsentry{CannyEdgeDetector}.
.3 \clsentry{GradientCalculator}; returns \texttt{Gradient}.
.2 \textbf{imageparam} --- image texture parameters (Chapter~\ref{chap:imageparam}).
.3 \clsentry{BaseParamCalculator}; uses \texttt{PatchSize}.
.3 \clsentry{MeanParamCalculator}.
.3 \clsentry{StdDeviationParamCalculator}.
.3 \clsentry{SkewnessParamCalculator}.
.3 \clsentry{KurtosisParamCalculator}.
.3 \clsentry{RelativeSmoothnessParamCalculator}.
.3 \clsentry{EntropyParamCalculator}.
.3 \clsentry{UniformityParamCalculator}.
.3 \clsentry{TContrastParamCalculator}.
.3 \clsentry{FractalDimParamCalculator}.
.3 \clsentry{TDirectionalityParamCalculator}; uses \texttt{PeakDetector}.
}
\end{minipage}

\medskip
\noindent\begin{minipage}{\textwidth}
\small
\DTsetlength{0.2em}{1em}{0.2em}{0.4pt}{1.6pt}
\dirtree{%
.1 \textbf{swdatatoolkit} (continued).
.2 \textbf{magfield} --- per-pixel field maps (Chapter~\ref{chap:magfield}).
.3 \clsentry{HorizontalFieldCalculator}.
.3 \clsentry{TotalFieldCalculator}.
.3 \clsentry{VerticalCurrentFieldCalculator}.
.3 \clsentry{GreenpotCalculator}.
.3 \clsentry{GreenFunctionPotentialCalculator}.
.3 \clsentry{LinearForceFreeFieldCalculator}.
.3 \clsentry{ShearAngleFieldCalculator}.
.2 \textbf{magparam} --- active-region parameters (Chapter~\ref{chap:magparam}).
.3 \clsentry{FluxCalculator}.
.3 \clsentry{GammaCalculator}.
.3 \clsentry{BaseDerivativeCalculator}.
.4 \clsentry{TotalDerivativeCalculator}.
.4 \clsentry{HorizontalDerivativeCalculator}.
.4 \clsentry{ZDerivativeCalculator}.
.4 \clsentry{LOSDerivativeCalculator}.
.3 \clsentry{TotalVerticalCurrentCalculator}.
.3 \clsentry{AlphaCalculator}.
.3 \clsentry{HelicityCalculator}.
.3 \clsentry{PolarityCalculator}.
.3 \clsentry{FreeEnergyCalculator}.
.3 \clsentry{ShearAngleCalculator}.
.3 \clsentry{LorentzCalculator}.
.3 \clsentry{RValueCalculator}.
.3 \clsentry{ShearWeightedRValueCalculator}.
.3 \clsentry{HelicityGuessCalculator}.
.3 \textbf{mct} --- magnetic charge topology and $B_\text{eff}$ (Section~\ref{sec:mct}).
.4 \clsentry{BeffCalculator}.
.4 \clsentry{SepLengthCalculator}; returns \texttt{SepLength}.
.4 \clsentry{MCTModelCreator}; returns \texttt{MCTModel}.
.4 \clsentry{MCTPartitioner}.
.4 \clsentry{ClusterPartitioner}.
.4 \clsentry{ClusterConnector}.
.4 \clsentry{DownhillClusterPartitioner}.
.4 \clsentry{IslandSwallower}.
.4 \clsentry{ClusterCorrector}.
.4 \clsentry{SmallFluxStructureRemover}.
.4 \clsentry{FluxSpreader}; returns \texttt{FluxDistributions}.
.4 \clsentry{GreedyOptimizer}; returns \texttt{PathMap}.
.2 \textbf{slicer} --- labelled sample windows (Chapter~\ref{chap:features}).
.3 \clsentry{HarpDataSlicer}.
.2 \textbf{features} --- time-series features (Chapter~\ref{chap:features}).
.3 \clsentry{StatisticalFeatures}.
.3 \clsentry{PySparkFeatureExtractor}.
}
\end{minipage}

%==========================================================
\chapter{Data Acquisition Capabilities}\label{chap:datasources}
%==========================================================

The \texttt{datasources} module contains the downloader classes and supporting enumerations used to retrieve solar images, active-region products, event tables, and related time-series data from external archives and services. The classes fall into three groups: those that retrieve image data (full-disk AIA images and HMI active-region patches) as FITS files on local disk (Section~\ref{sec:img_retrieval}); those that retrieve and normalize catalogs of discrete solar events such as flares, active regions, and CMEs (Section~\ref{sec:event_downloaders}); and those that retrieve continuously sampled in-situ or irradiance measurements (Section~\ref{sec:ts_downloaders}). Every class accepts an optional \texttt{logging.Logger} at construction; if none is supplied, messages are written to \texttt{stderr}. Network access is wrapped in bounded retries, so a transient failure on one request does not abort a long multi-day query. \href{https://dmlab.cs.gsu.edu/docs/spaceweather_toolkit/}{\textbf{swdatatoolkit}} therefore provides a class-based entry point for assembling heterogeneous space weather datasets in a reproducible way.

%////////////////////Section\\\\\\\\\\\\\\\\\\\\\\\\\\\\\\
\section{Solar Image and Active-Region Retrieval}\label{sec:img_retrieval}

Both image downloaders retrieve data from the Joint Science Operations Center (JSOC) at Stanford through its export system. Because a JSOC export is an asynchronous, e-mail-notified request, both constructors require an e-mail address registered with JSOC. Both write the requested FITS files to a local directory and return the locations of what was downloaded, rather than holding image data in memory.

%---------------------------------------------------------
\subsection{\texttt{AIADownloader}}\classlabel{AIADownloader}

\paragraph{What it does.} \texttt{AIADownloader} downloads level-1 full-disk images from the Atmospheric Imaging Assembly (AIA) on board SDO~\cite{lemen:sdoAIA}. A request is described by a time range, a single wavelength channel, and a cadence (a \texttt{datetime.timedelta}); the first record is taken at the start of the range and further records are taken at the requested cadence. The workflow can be done in one step with \texttt{download\_records\_for\_range()}, or in two steps: \texttt{get\_records\_for\_range()} runs the query and returns the matching record list without downloading anything, so it can be inspected or filtered, and \texttt{download\_records\_for\_response()} then downloads exactly those records.

\paragraph{Where it gets the data.} The query is made against JSOC through SunPy's JSOC client, requesting only the \texttt{image} segment. The seven EUV channels (94, 131, 171, 193, 211, 304, and 335~\AA) come from the \texttt{aia.lev1\_euv\_12s} series, and the two UV channels (1600 and 1700~\AA) come from \texttt{aia.lev1\_uv\_24s}. The channel is chosen with the \texttt{WaveBand} enumeration, which has one member per channel (\texttt{Wave\_94}, \texttt{Wave\_131}, \dots, \texttt{Wave\_1700}). The same enumeration is used by the cutout processor (Section~\ref{cls:AIACutOutProcessor}) to pick the emission height for a channel.

\paragraph{How the data are stored and returned.} JSOC delivers the export as a tar archive. The downloader streams it to a partial file, renames it when the transfer completes, and extracts the FITS files into the data directory given to the constructor. If no directory is given, a temporary directory is created; it is deleted when the process exits, so files must be moved out of it before then. Both download methods return a list of paths to the FITS files that were downloaded. \texttt{get\_records\_for\_range()} returns the query result (a SunPy \texttt{JSOCResponse}), or \texttt{None} if the query failed.

%---------------------------------------------------------
\subsection{\texttt{HARPDownloader}}\classlabel{HARPDownloader}

\paragraph{What it does.} \texttt{HARPDownloader} retrieves Space-weather HMI Active Region Patch (SHARP) data~\cite{bobra2014helioseismic, Hoeksema2014}. These are cutouts of the HMI vector magnetogram and related products, made around each automatically tracked HMI Active Region Patch (HARP). Data can be requested in three ways: by HARP number for the whole life of the region (\texttt{download\_records\_for\_harp\_number()}), by HARP number restricted to a time window (\texttt{download\_records\_for\_harp\_number\_and\_range()}), or for every HARP observed in a time window (\texttt{download\_records\_for\_range()}). Two helper methods support incremental updates. \texttt{get\_max\_harp\_num()} returns the most recent HARP number at JSOC. \texttt{get\_harp\_process\_dates()} returns the date on which JSOC processed each record of a HARP, which can be compared with local copies to find records that have been reprocessed.

\paragraph{Where it gets the data.} Records come from the cylindrical-equal-area (CEA) SHARP series at JSOC. The \texttt{HARPDataType} enumeration selects either the definitive, science-quality series (\texttt{DEFINITIVE}, \texttt{hmi.sharp\_cea\_720s}) or the near-real-time series (\texttt{NRT}, \texttt{hmi.sharp\_cea\_720s\_nrt}, the default). The \texttt{filetypes} argument restricts the download to a subset of the data segments: the vector-field components \texttt{Br}, \texttt{Bt}, \texttt{Bp} and their errors \texttt{Br\_err}, \texttt{Bt\_err}, \texttt{Bp\_err}; the HARP \texttt{bitmap}; the disambiguation confidence \texttt{conf\_disambig}; and the line-of-sight \texttt{magnetogram}, \texttt{Dopplergram}, and \texttt{continuum} images. All segments are selected when the argument is omitted. Some NRT records are listed in the series but are not actually online. Requesting one of them makes the whole export fail, so before each export the downloader asks JSOC's \texttt{jsoc\_info} service which records are online and requests only those. A HARP with many records is split into partitions, each partition is exported separately, and a failed partition is retried up to five times without repeating partitions that already succeeded.

\paragraph{How the data are stored and returned.} Each HARP is written to its own subdirectory of the data directory, named by HARP number. Each partition's tar archive is extracted there, and the archive and the auxiliary files JSOC includes with the export (\texttt{.txt}, \texttt{.html}, \texttt{.json}, and so on) are deleted, leaving one FITS file per segment per time step. If the data already exist on disk, no new request is made. The temporary-directory behavior is the same as for \texttt{AIADownloader}. The download methods return a tuple of (i) the list of HARP numbers that downloaded successfully and (ii) the root directory that contains the per-HARP subdirectories. \texttt{get\_harp\_process\_dates()} returns a \texttt{pandas} \texttt{DataFrame} of processing dates, one per record.

%////////////////////Section\\\\\\\\\\\\\\\\\\\\\\\\\\\\\\
\section{Event and Catalog Downloaders}\label{sec:event_downloaders}

The event downloaders retrieve catalogs of discrete solar events and return them as \texttt{pandas} \texttt{DataFrame} objects with a fixed set of columns. An empty or failed query returns an empty frame with the same columns rather than raising an exception, so the results of several catalogs can be concatenated and cross-matched without special cases. All times are returned as timezone-naive UTC \texttt{datetime} values, and positions are heliographic latitude and longitude in degrees. Nothing is written to disk, and caching or persistence is left to the caller.

%---------------------------------------------------------
\subsection{\texttt{AIAFlareEventDownloader}}\classlabel{AIAFlareEventDownloader}

\paragraph{What it does.} \texttt{download\_data(start\_date, end\_date)} returns the flare events reported by SDO/AIA feature-recognition modules over a date range. These include flare positions measured in EUV images, which locate the source region more precisely than the GOES report can.

\paragraph{Where it gets the data.} Events are queried from the Heliophysics Event Knowledgebase (HEK) through SunPy's HEK client, using event type \texttt{FL} and observatory \texttt{SDO}. Long ranges are split into 30-day batches, and each batch is retried up to \texttt{max\_retries} times (default 3) with a delay of \texttt{retry\_delay} seconds between attempts. A batch that still fails is logged and skipped, so the remaining batches are still returned. For windows before SDO was launched, the observatory filter is dropped and generic HEK flare events are returned, so historical queries still produce results.

\paragraph{How the data are returned.} The method returns a \texttt{DataFrame} with columns \texttt{time\_tag} (HEK archive date), \texttt{begin\_time}, \texttt{begin\_class}, \texttt{max\_time}, \texttt{max\_class} (GOES class), \texttt{end\_time}, \texttt{end\_class}, \texttt{satellite}, \texttt{NOAA\_AR} (a nullable integer that is missing when no region was associated), \texttt{latitude}, and \texttt{longitude}. Rows are de-duplicated on \texttt{begin\_time} and sorted by \texttt{begin\_time} in descending order. The columns are identical to those of \texttt{GOESFlareEventDownloader}, so the two catalogs can be compared directly.

%---------------------------------------------------------
\subsection{\texttt{GOESFlareEventDownloader}}\classlabel{GOESFlareEventDownloader}

\paragraph{What it does.} \texttt{download\_data(start\_date, end\_date)} returns the GOES X-ray flare catalog for a date range. This is the standard list of flare start, peak, and end times and GOES classes used to label flare-forecasting datasets.

\paragraph{Where it gets the data.} Archival events come from HEK (event type \texttt{FL}, observatory \texttt{GOES}), queried in 30-day batches with bounded retries. HEK can lag real time by several days, so when the requested window overlaps the most recent seven days, the downloader also reads NOAA SWPC's seven-day X-ray flare feed (\texttt{xray-flares-7-day.json}) and merges it with the HEK results.

\paragraph{How the data are returned.} The output columns are the same as those of the AIA flare downloader. Rows are filtered to peak times (\texttt{max\_time}) inside the requested window, and an event reported by both sources appears only once (duplicates are removed on \texttt{max\_time}). Near-real-time and archival queries therefore return the same columns.

%---------------------------------------------------------
\subsection{\texttt{SRSDownloader}}\classlabel{SRSDownloader}

\paragraph{What it does.} \texttt{fetch\_data(start\_date, end\_date)} parses NOAA's daily Solar Region Summary (SRS). The SRS is the official daily list of numbered active regions with their positions and sunspot-group properties, and it is used to associate HARPs and flare reports with NOAA region numbers.

\paragraph{Where it gets the data.} One plain-text SRS file per day is read from the NOAA NCEI archive of SWPC daily reports (\texttt{solar\_region\_summaries}). Both of the report's sections are parsed: regions with sunspots, and H-alpha plages without spots (section IA).

\paragraph{How the data are returned.} The method returns a \texttt{DataFrame} with one row per region per day and columns \texttt{date\_tag}, \texttt{region\_number}, \texttt{latitude}, \texttt{longitude}, \texttt{area} (in millionths of a hemisphere), \texttt{z\_class} (the McIntosh classification, or \texttt{Plage} for spotless regions), \texttt{ll} (longitudinal extent), and \texttt{num\_spots}. Before 2002, NOAA region numbers were reported with four digits. These are moved onto the modern five-digit numbering by adding 10{,}000, so that region numbers from different epochs refer to the same region and can be joined with other catalogs.

%---------------------------------------------------------
\subsection{\texttt{SSWEventDownloader}}\classlabel{SSWEventDownloader}

\paragraph{What it does.} \texttt{scrape(start, end)} collects the flare reports published by the SolarSoft (SSW) ``Latest Events'' service of the Lockheed Martin Solar and Astrophysics Laboratory. These reports give a flare location, estimated from imaging data, for many events that have no NOAA region assignment.

\paragraph{Where it gets the data.} The data come from the SSW latest-events archive (\url{https://www.lmsal.com/solarsoft/latest_events_archive.html}). The scraper reads the archive's landing page and finds the daily \texttt{last\_events\_YYYY\dots} index pages that fall in the requested range. It then follows each index page to the individual flare pages (\texttt{gev\_*.html}) and parses the flare tables on them. Each page fetch is retried with a timeout.

\paragraph{How the data are returned.} The method returns a \texttt{DataFrame} with columns \texttt{goes\_class}, \texttt{goes\_location\_lat}, \texttt{goes\_location\_lon}, \texttt{noaa\_active\_region}, \texttt{start\_time}, \texttt{peak\_time}, and \texttt{end\_time}. Pages that could not be parsed are tracked separately, so that gaps in the scraped catalog can be found.

%---------------------------------------------------------
\subsection{\texttt{CMEDownloader}}\classlabel{CMEDownloader}

\paragraph{What it does.} \texttt{get\_reports(start\_time, end\_time, report, category)} retrieves coronal mass ejection detections from the CACTus catalog. CACTus detects CMEs automatically in SOHO/LASCO coronagraph images. The \texttt{report} argument selects either CMEs (\texttt{"CME"}) or the smaller outflows CACTus also reports (\texttt{"FLOW"}). The \texttt{category} argument selects either the quick-look catalog (\texttt{"QUICKLOOK"}, the default) or the level-zero catalog (\texttt{"LEVEL\_ZERO"}). The allowed values are the member names of the \texttt{CACTUSReportType} and \texttt{ReportCat} enumerations, and any other value raises a \texttt{ValueError}.

\paragraph{Where it gets the data.} The monthly \texttt{cmecat.txt} bulletins of the CACTus LASCO catalog are read from the Royal Observatory of Belgium's Solar Influences Data Analysis Center (SIDC), one bulletin per month in the requested range. Monthly bulletins can lag behind current detections, so the most recently issued bulletin (\texttt{cactus/out/cmecat.txt}) is also read. For each day, any of its detections later than the latest detection already in the monthly catalog are appended.

\paragraph{How the data are returned.} The method returns a \texttt{DataFrame} with one row per detection. Its columns come from the CACTus table header and include the onset time \texttt{t}, duration, principal angle, angular width, median velocity and its spread, minimum and maximum velocities, and the halo flag, together with the \texttt{Issue Date} of the bulletin the row came from. Non-time columns are converted to numeric types.

%---------------------------------------------------------
\subsection{\texttt{XRTHinodeEventDownloader}}\classlabel{XRTHinodeEventDownloader}

\paragraph{What it does.} \texttt{get\_xrt\_events\_between(start\_time, end\_time)} returns the flares in the Hinode X-Ray Telescope (XRT) flare catalog whose peak time falls inside the requested window, with both ends of the window included. XRT soft X-ray imaging gives an independent estimate of flare location that can be used to check GOES and AIA location assignments.

\paragraph{Where it gets the data.} The full catalog is downloaded as a single CSV file from the Harvard--Smithsonian Center for Astrophysics (\texttt{xrt.cfa.harvard.edu/flare\_catalog/xrt\_flarecat.csv}). The request uses the \texttt{timeout} and \texttt{max\_retries} set in the constructor, and the time filter is applied locally.

\paragraph{How the data are returned.} The method returns a \texttt{DataFrame} with columns \texttt{start}, \texttt{end}, \texttt{peak}, \texttt{class} (GOES class), \texttt{lat}, \texttt{lon}, and \texttt{region} (NOAA region as a nullable integer).

%////////////////////Section\\\\\\\\\\\\\\\\\\\\\\\\\\\\\\
\section{Time-Series Downloaders}\label{sec:ts_downloaders}

The time-series downloaders retrieve continuously sampled measurements rather than discrete events. Each one takes a start and end \texttt{datetime} (and, for neutron monitors, a station name). Each splits the request into the daily, monthly, or yearly files of its source archive, downloads and parses those files, joins the pieces, and trims the result to the requested window. Each returns a \texttt{DataFrame} indexed by observation time, so that products from different sources can be aligned with \texttt{pandas} joins or resampling. Each class also has a lower date limit set by the coverage of its source, and a query that starts earlier raises a \texttt{ValueError} instead of silently returning partial data. Nothing is cached on disk.

%---------------------------------------------------------
\subsection{\texttt{SXRDownloader}}\classlabel{SXRDownloader}

\paragraph{What it does.} \texttt{get\_one\_minute\_avg\_data(start\_time, end\_time, satellite=None)} returns one-minute-averaged soft X-ray irradiance from the GOES X-Ray Sensor (XRS) in its two channels: \texttt{xrsa} (0.05--0.4~nm) and \texttt{xrsb} (0.1--0.8~nm; the GOES flare class is defined from this channel). Several GOES satellites usually operate at once, and the satellite NOAA designates as primary has changed many times since 1995. The downloader keeps a table of the primary satellite for each period and, for each day of the query, downloads data from the satellite that was primary on that day. A specific satellite (GOES-8 or later) can be requested instead, for example to fill gaps in the primary record from a secondary spacecraft. The static helpers \texttt{has\_archived\_data()} and \texttt{archived\_satellites()} report, for a given day, which satellites have 1-minute science data in the NOAA archive. Days with no archived data for the requested satellite are skipped without making a request.

\paragraph{Where it gets the data.} Archival data are the NOAA NCEI XRS level-2 one-minute-average science products (\texttt{xrsf-l2-avg1m\_science}), stored as daily NetCDF files. Files for GOES-8 through GOES-15 are read from the NCEI \texttt{goes-space-environment-monitor} archive, and files for the GOES-R series (GOES-16 onward) are read from the NGDC platform archive. The two archives use slightly different variable names, and both layouts are handled. For the most recent days, the NOAA SWPC one-day and seven-day JSON feeds are used: the primary feed, or the secondary feed when the requested satellite is not primary on that day.

\paragraph{How the data are returned.} The method returns a time-indexed \texttt{DataFrame} with columns \texttt{xrsa\_flux} and \texttt{xrsb\_flux} (W\,m$^{-2}$), quality flags \texttt{xrsa\_flag(s)} and \texttt{xrsb\_flag(s)} (0 marks good data; nonzero values mark eclipse, contamination, and similar conditions), and \texttt{xrsa\_num} and \texttt{xrsb\_num}, the number of samples in each one-minute average. In the older archives, missing flux is stored as $-9999$. Queries before 1 March 1995 raise an exception.

%---------------------------------------------------------
\subsection{\texttt{ProtonDownloader}}\classlabel{ProtonDownloader}

\paragraph{What it does.} \texttt{get\_five\_min\_avg\_data(start\_time, end\_time)} returns five-minute integral proton fluxes from the primary GOES satellite above the thresholds $\geq$1, $\geq$5, $\geq$10, $\geq$30, $\geq$50, $\geq$60, $\geq$100, and $\geq$500~MeV. The $\geq$10~MeV channel defines NOAA's solar energetic particle (SEP) event threshold.

\paragraph{Where it gets the data.} Archival fluxes come from the NASA CCMC Integrated Space Weather Analysis (ISWA) system through its HAPI interface. For the most recent days, they come from the NOAA SWPC one-day and seven-day primary-satellite integral-proton JSON feeds. The query is made day by day and the days are joined. Coverage begins with GOES-14 on 1 April 2010, and earlier queries raise an exception.

\paragraph{How the data are returned.} The method returns a time-indexed \texttt{DataFrame} with one column per energy threshold (\texttt{>=1 MeV}, \dots, \texttt{>=500 MeV}), in particle flux units (pfu, protons\,cm$^{-2}$\,s$^{-1}$\,sr$^{-1}$).

%---------------------------------------------------------
\subsection{\texttt{OMNIWebDownloader}}\classlabel{OMNIWebDownloader}

\paragraph{What it does.} \texttt{get\_five\_min\_avg\_data(start\_time, end\_time)} returns the high-resolution five-minute OMNI data set. This is a multi-spacecraft compilation of near-Earth solar-wind plasma and interplanetary magnetic field (IMF) measurements, time-shifted to the Earth's bow shock, together with geomagnetic indices and the GOES proton fluxes above 10, 30, and 60~MeV.

\paragraph{Where it gets the data.} The data come from NASA Goddard Space Flight Center's OMNIWeb service. The downloader submits a form request to the OMNIWeb CGI interface for the \texttt{omni\_5min\_def} data set, asking for all of variables 4 through 48. The plain-text table in the HTML response is then parsed. Queries before 1 January 1982 raise an exception.

\paragraph{How the data are returned.} The method returns a time-indexed \texttt{DataFrame} whose column names are the OMNIWeb variable descriptions. The available fields are described in the high-resolution OMNI format document. The method returns \texttt{None} if the request fails.

%---------------------------------------------------------
\subsection{\texttt{SOHOEPHINDownloader}}\classlabel{SOHOEPHINDownloader}

\paragraph{What it does.} \texttt{get\_one\_minute\_avg\_data(start\_time, end\_time)} returns one-minute-averaged energetic-particle intensities measured by the Electron Proton Helium Instrument (EPHIN) on SOHO. EPHIN observes from the L1 point and measures electrons, protons, and helium nuclei, which makes it useful for SEP studies.

\paragraph{Where it gets the data.} The data are the level-3 one-minute files from the COSTEP archive at Christian-Albrechts-Universit\"at zu Kiel, one file per day. Coverage begins in 1996.

\paragraph{How the data are returned.} The method returns a \texttt{DataFrame} with a \texttt{date} column followed by the instrument \texttt{status} and the intensity channels listed in the file header (see the EPHIN level-3 documentation for channel definitions). It returns \texttt{None} if nothing could be retrieved.

%---------------------------------------------------------
\subsection{\texttt{NeutronDownloader}}\classlabel{NeutronDownloader}

\paragraph{What it does.} \texttt{get\_one\_min\_avg\_data(start\_time, end\_time, monitor="OULU")} returns one-minute ground-level neutron-monitor data. These record the secondary neutrons produced in the atmosphere by galactic cosmic rays and, during ground-level enhancements, by the most energetic solar protons. Three stations are supported: \texttt{OULU} (Oulu, Finland, available from 1 April 2010), and \texttt{DOMC} and \texttt{DOMB} (the two Concordia/Dome~C monitors in Antarctica, available from 12 March 2016). Any other station name raises a \texttt{ValueError}.

\paragraph{Where it gets the data.} Each day is requested separately from the online query service of the University of Oulu Cosmic Ray Station.

\paragraph{How the data are returned.} The method returns a time-indexed \texttt{DataFrame} with columns \texttt{uncorr} (raw count rate, counts\,min$^{-1}$), \texttt{corr} (count rate corrected for barometric pressure and detector efficiency), and \texttt{pressure} (mbar). It returns \texttt{None} if nothing was retrieved.

%---------------------------------------------------------
\subsection{\texttt{DSCOVRDownloader}}\classlabel{DSCOVRDownloader}

\paragraph{What it does.} \texttt{get\_fc\_data(start\_time, end\_time)} returns one-minute solar-wind plasma measurements from the Faraday Cup (FC) instrument on the Deep Space Climate Observatory (DSCOVR). DSCOVR is NOAA's operational L1 solar-wind monitor.

\paragraph{Where it gets the data.} The data are the archived daily FC files in NOAA NCEI's DSCOVR archive, organized by year and month. Coverage begins on 1 August 2016, and earlier queries raise an exception.

\paragraph{How the data are returned.} The method returns a \texttt{DataFrame} of one-minute records. The columns are the proton and alpha-particle speed, density, and temperature; the bulk velocity components in GSE and GSM coordinates; the number of samples in each average; and the full set of FC quality flags (fill, calibration-mode, maneuver, out-of-range, peak-confusion, and per-quantity ``bad'' flags), so that users can apply their own quality screening.

%==========================================================
\chapter{Image Processing and Structural Analysis}\label{chap:imageproc}
%==========================================================

Solar imagery is often analyzed not only in terms of raw intensity values but also through structural descriptors that characterize morphology, texture, and boundary information.
\href{https://dmlab.cs.gsu.edu/docs/spaceweather_toolkit/}{\textbf{swdatatoolkit}} includes components for image preprocessing and edge detection to support these needs.

%////////////////////Section\\\\\\\\\\\\\\\\\\\\\\\\\\\\\\
\section{Image Preprocessing}

The image processing utilities support operations such as image subsetting and region extraction. These functions are especially relevant when studies focus on active regions rather than full-disk solar images. Preprocessing helps standardize inputs prior to feature extraction, classification, or comparison across observations.

%////////////////////Section\\\\\\\\\\\\\\\\\\\\\\\\\\\\\\
\section{Edge Detection}

The edge-detection components provide tools for identifying structural boundaries and intensity discontinuities within images. In the context of solar physics, such methods may be used to characterize spatial features, support segmentation, or provide auxiliary descriptors for texture analysis.

%////////////////////Section\\\\\\\\\\\\\\\\\\\\\\\\\\\\\\
\section{Research Utility}

These image-analysis tools are useful in several contexts:

\begin{itemize}
    \item Active-region morphology studies
    \item Preprocessing for machine learning models
    \item Boundary and contour analysis
    \item Extraction of shape-based descriptors
\end{itemize}

%////////////////////Section\\\\\\\\\\\\\\\\\\\\\\\\\\\\\\
\section{\texttt{AIACutOutProcessor} in \texttt{imageproc}}\classlabel{AIACutOutProcessor}\label{sec:aia_cutout}

The \texttt{imageproc} module contains one class, \texttt{AIACutOutProcessor}. It extracts from a full-disk AIA image the region that corresponds to an active region whose extent is given in heliographic (Carrington) coordinates, normally the bounding box of a HARP. The resulting cutouts pair each SHARP magnetogram with EUV and UV images of the same region, and the image parameters of Chapter~\ref{chap:imageparam} are computed over them. The procedure below is the one used to build the AIA portion of the SWAN-SF-V2 dataset.

\paragraph{Inputs.} The method \texttt{process\_cutout()} takes a full-disk AIA image as a SunPy map, the Carrington longitude and latitude of the lower-left and upper-right corners of the region (in degrees), the observation time, and the \texttt{WaveBand} of the image. A HARP's bounding box is usually given by the minimum bounding rectangle of the pixels in its bitmap that are used in the SHARP parameter calculations. Images whose \texttt{QUALITY} header keyword is nonzero are rejected, as are boxes of zero width or height; in both cases the method returns \texttt{None}.

\paragraph{Step 1: proportional dilation.} The HARP bounding box is first enlarged by a fraction of its size, set by the \texttt{extension} constructor argument (default 0.2, a 20\% expansion). Half of the expansion is added to each side, and the amount added to each side is capped at \texttt{max\_extension\_as} arcseconds (default 25\arcsec), converted to heliographic degrees using the apparent solar radius \textsc{rsun\_obs}. This captures loop footpoints and surrounding emission just outside the magnetogram patch without letting large regions grow without limit. Boxes that cross the 0\textdegree/360\textdegree\ Carrington meridian are handled.

\begin{figure}[htb!]
    \centering
    \includegraphics[width=\textwidth]{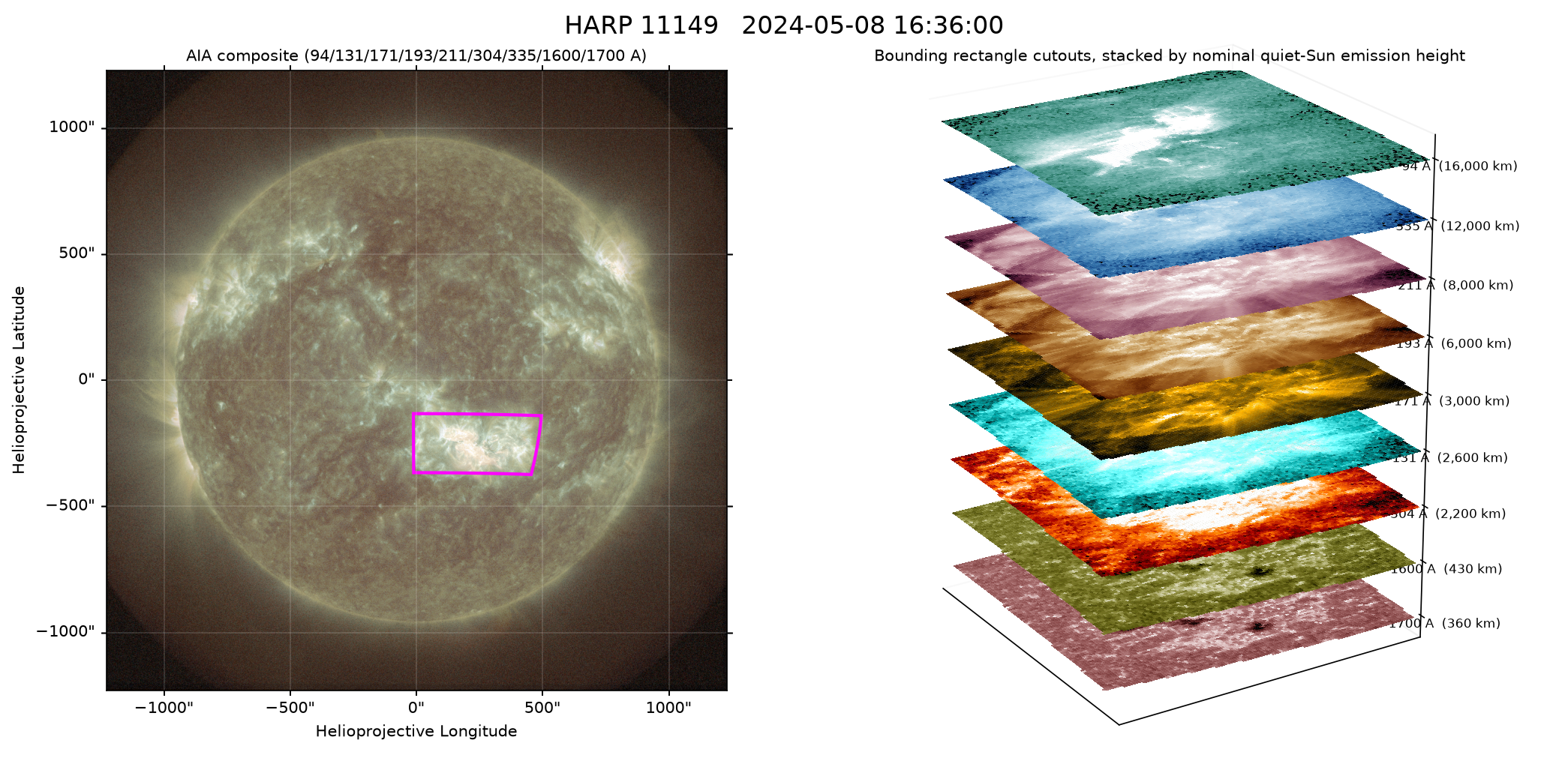}
    \caption{AIA cutouts produced by \texttt{AIACutOutProcessor} for HARP~11149 at 2024-05-08 16:36:00. Left: the extended cutout region (magenta) drawn on a nine-channel AIA composite; the curved edges come from projecting the Carrington-coordinate box onto the image plane. Right: the cutout from each channel, stacked by nominal quiet-Sun emission height.}
    \label{fig:cutout_visual}
\end{figure}

\paragraph{Step 2: projection onto the AIA image.} The dilated box is sampled on a regular Carrington longitude--latitude grid whose step corresponds to one AIA pixel at disk center (\(\arcsin(\textsc{cdelt1}/\textsc{rsun\_obs})\)). The grid is converted to helioprojective coordinates as seen from Earth at \texttt{obs\_time}, and then to AIA pixel indices using the image's reference pixel and plate scale. Marking these pixels produces a mask on the full-disk image. Because foreshortening is applied point by point, the mask's outline is a curved quadrilateral rather than a rectangle (Figure~\ref{fig:cutout_visual}, left).

\paragraph{Step 3: emission-height extension.} Coronal emission is formed above the photosphere, and away from disk center its projected position is displaced toward the limb. A cutout that only follows the photospheric footprint therefore misses part of the loops rooted in the region. Following \cite{dissauer_properties_2023}, the mask is extended on the two edges that face the nearest limb, by an amount that grows with distance from disk center. The emission height \(h\) depends on the channel: 2.5~Mm for 1600 and 1700~\AA, which form near the temperature minimum and low chromosphere~\cite{sanjay_formation_2024}, and 75~Mm for all EUV channels. For a region centered at longitude \(\phi\) (measured from the central meridian, \textsc{crln\_obs}) and latitude \(\theta\), the outward radial unit vector is \(\hat r=(\sin\phi\cos\theta,\ \sin\theta,\ \cos\phi\cos\theta)\), with \(\hat z\) along the line of sight. Raising the emission by \(h\) along \(\hat r\) shifts its apparent position by the tangential part of \(h\hat r\):
\begin{equation}
    \Delta_\lambda = h\,|\sin\phi\cos\theta|, \qquad
    \Delta_\beta = h\,|\sin\theta|, \qquad
    \sqrt{\Delta_\lambda^2+\Delta_\beta^2} = h\sin\psi,\quad \psi=\arccos(\cos\phi\cos\theta).
    \label{eq:emission_offset}
\end{equation}
The offsets are converted to pixel margins \(n_\lambda=\lceil\Delta_\lambda/m\rceil\) and \(n_\beta=\lceil\Delta_\beta/m\rceil\), where \(m=\textsc{cdelt1}\cdot\textsc{rsun\_ref}/\textsc{rsun\_obs}\) is the number of meters per pixel at disk center. The implementation checks at run time that the two components reproduce the magnitude \(h\sin\psi\). The signs of \(\phi\) and \(\theta\) give the quadrant of the disk the region lies in, and that determines which pair of edges (east or west, north or south) is extended. The extension is applied row by row and column by column from the existing mask boundary, so the curved outline of the mask is preserved. The geometry is illustrated in Figure~\ref{fig:cutextend}.

\paragraph{Step 4: cropping and output.} The full-disk image is multiplied by the extended mask and cropped to the minimum bounding rectangle of the mask. The method returns a SunPy \texttt{GenericMap} holding the cropped intensity array, with pixels outside the mask set to \texttt{NaN}. The map carries the header of the full-disk image, and its reference-pixel keywords are not shifted to the cutout. The texture parameters of Chapter~\ref{chap:imageparam} ignore NaNs, so only pixels inside the mask contribute to them.

Figure~\ref{fig:cutout_visual} shows the result for HARP~11149 (NOAA AR~13664) at 2024-05-08 16:36. The left panel outlines the dilated and extended region on a composite of all nine channels. The right panel shows the nine resulting cutouts, stacked by the nominal quiet-Sun formation height of each channel.

% Adapted from aia-mvts/documentation/fig/aia_cutout_emission_height.tex (SWAN-SF-V2 document).
% The note on the superseded per-axis split was removed: AIACutOutProcessor._emission_offset_meters now
% implements the direction-cosine split shown here.
\begin{figure}[htb!]
    \centering

    \noindent
    \begin{tikzpicture}[
      >={Stealth[length=2.3mm]},
      every node/.style={font=\small, text=ink},
      lbl/.style={font=\footnotesize},
      tny/.style={font=\scriptsize},
      guide/.style={densely dashed, gray!65, line width=0.35pt},
      trim left,
    ]
    
    % =====================================================================
    % PANEL (a): side view, looking down on the observer-Sun plane
    % =====================================================================
    \begin{scope}[xshift=3.7cm]
      \def\R{2.25}       % solar radius, drawing units
      \def\Rh{3.10}      % R + h  (h exaggerated for legibility)
      \def\pa{50}        % heliocentric angle psi of the region
    
      % emission shell and photosphere
      \draw[sunedge, densely dotted, line width=0.6pt] (0,0) circle (\Rh);
      \filldraw[fill=sunfill, draw=sunedge, line width=0.8pt] (0,0) circle (\R);
      \node[tny, text=sunedge, anchor=east] at (-0.18,{\Rh+0.20})
           {emission shell, radius $R_\odot+h$};
    
      % plane of sky (perpendicular to the line of sight, through Sun centre)
      \draw[ink, line width=0.7pt] (0,-2.65) -- (0,3.60);
      \node[lbl, rotate=90, anchor=south] at (-0.14,-2.00) {plane of sky};
    
      % line of sight through disk centre
      \draw[guide] (0,0) -- (3.65,0);
      \draw[->, ink, line width=0.7pt] (2.90,0) -- (3.63,0);
      \node[tny, anchor=south east, fill=white, inner sep=1pt]
           at (3.65,0.13) {to observer};
      \fill[ink] (0,0) circle (1.3pt) node[lbl, below left=-1pt] {$O$};
      \fill[ink] (\R,0) circle (1.3pt);
      \node[tny, below=2pt, fill=white, inner sep=1pt]
           at (\R,0) {sub-observer point};
    
      % the region of interest on the photosphere, at heliocentric angle psi
      \coordinate (P)  at ({\R*cos(\pa)},  {\R*sin(\pa)});
      \coordinate (Pp) at ({\Rh*cos(\pa)}, {\Rh*sin(\pa)});
      \draw[guide] (0,0) -- (Pp);
      \fill[roi] (P) circle (1.8pt);
      \node[tny, text=roi, anchor=north west, fill=white, inner sep=1pt]
           at ($(P)+(0.10,-0.07)$) {region at $(\phi,\theta)$};
    
      % radial emission height h
      \draw[->, ext, line width=1.1pt] (P) -- (Pp);
      \node[lbl, text=ext, anchor=east] at ($(P)!0.55!(Pp)+(-0.05,0.02)$) {$h$};
    
      % heliocentric angle psi at the Sun centre ...
      \draw[ink, line width=0.5pt]
            (0.80,0) arc[start angle=0, end angle=\pa, radius=0.80];
      \node[lbl] at ({1.00*cos(0.5*\pa)},{1.00*sin(0.5*\pa)}) {$\psi$};
      % ... reappears between the local radial and the line of sight
      \draw[guide] (P) -- ++(1.10,0);
      \draw[ink, line width=0.5pt]
            ($(P)+(0.47,0)$) arc[start angle=0, end angle=\pa, radius=0.47];
      \node[tny] at ($(P)+({0.64*cos(0.5*\pa)},{0.64*sin(0.5*\pa)})$) {$\psi$};
    
      % projection of the footpoint and of the top of the emission column
      \draw[guide] (P)  -- (0,{\R*sin(\pa)});
      \draw[guide] (Pp) -- (0,{\Rh*sin(\pa)});
      \fill[roi] (0,{\R*sin(\pa)})  circle (1.5pt);
      \fill[ext] (0,{\Rh*sin(\pa)}) circle (1.5pt);
    
      % brace marking the apparent (plane-of-sky) offset rho = h sin(psi)
      \draw[ext, decorate, decoration={brace, amplitude=4pt}, line width=0.7pt]
            (-0.33,{\Rh*sin(\pa)}) -- (-0.33,{\R*sin(\pa)});
      \node[tny, text=ext, anchor=east] at (-0.50,{0.5*(\Rh+\R)*sin(\pa)})
           {$\rho = h\sin\psi$};
    
      \node[font=\bfseries\small, anchor=north west] at (-3.75,-3.35)
           {(a) the magnitude of the offset};
      \node[tny, anchor=north west, text width=7.5cm, align=left]at (-3.75,-3.90)
           {A radially outward emission column of height $h$ makes an angle
            $\psi=\arccos(\cos\phi\cos\theta)$ with the line of sight, so only
            its tangential part is visible and the region's apparent position
            shifts by $\rho=h\sin\psi$ --- zero at the disk centre, growing to
            the full $h$ at the limb. };
    \end{scope}
    
    % =====================================================================
    % PANEL (b): plane-of-sky view -- splitting rho between the two axes
    % =====================================================================
    \begin{scope}[xshift=12.2cm]
      \def\Rd{2.45}
      % Apparent position of the region centre is Rd*(sin(phi)cos(theta),
      % sin(theta)); drawn for phi = 22 deg, theta = 16 deg.  The offset
      % h*(sin(phi)cos(theta), sin(theta)) is PARALLEL to that vector, so the
      % arrow below lies exactly on the ray from the disk centre.
      \def\cx{0.883}\def\cy{0.676}
      \def\xa{0.467}\def\xb{1.299}    % dilated-ROI bounding box in x
      \def\ya{0.356}\def\yb{0.996}    % dilated-ROI bounding box in y
      \def\dl{0.639}\def\db{0.489}    % offset components (drawn enlarged)
    
      \filldraw[fill=sunfill, draw=sunedge, line width=0.8pt] (0,0) circle (\Rd);
      \draw[guide] (-\Rd,0) -- (\Rd,0);
      \draw[guide] (0,-\Rd) -- (0,\Rd);
      \node[tny, anchor=west]  at ({\Rd+0.05},0) {W};
      \node[tny, anchor=south] at (0,{\Rd+0.03}) {N};
    
      % the ray from the disk centre through the region: the offset direction
      \draw[guide] (0,0) -- (2.042,1.565);
      \node[tny, text=gray, anchor=west, rotate=37.4] at (2.07,1.59){radially outward};
      \fill[ink] (0,0) circle (1.3pt);
      \node[tny, below left=-1pt, fill=white, inner sep=1pt] at (0,0) {disk centre};
    
      % margin, added on the two limb-facing edges only
      \fill[ext, opacity=0.16] (\xb,\ya) rectangle ({\xb+\dl},{\yb+\db});
      \fill[ext, opacity=0.16] (\xa,\yb) rectangle ({\xb+\dl},{\yb+\db});
      \draw[ext, densely dashed, line width=0.8pt]
            (\xa,\ya) rectangle ({\xb+\dl},{\yb+\db});
      \draw[roi, line width=1.0pt] (\xa,\ya) rectangle (\xb,\yb);
    
      % the offset vector and its decomposition: rho is the hypotenuse
      \draw[ext, line width=0.6pt] (\cx,\cy) -- ({\cx+\dl},\cy)
                                             -- ({\cx+\dl},{\cy+\db});
      \draw[->, ext, line width=1.1pt] (\cx,\cy) -- ({\cx+\dl},{\cy+\db});
      \draw[ink, line width=0.45pt]
            ({\cx+\dl-0.11},\cy) -- ({\cx+\dl-0.11},{\cy+0.11})
                                 -- ({\cx+\dl},{\cy+0.11});
      \fill[roi] (\cx,\cy) circle (1.7pt);
      \node[tny, text=roi, anchor=north east, xshift=1pt, yshift=1pt]
           at (\cx,\cy) {$(\phi,\theta)$};
      \node[tny, text=ext, anchor=north] at ({\cx+0.5*\dl},{\cy-0.03})
           {$\Delta_{\lambda}$};
      \node[tny, text=ext, anchor=west]  at ({\cx+\dl+0.05},{\cy+0.5*\db})
           {$\Delta_{\beta}$};
      \node[tny, text=ext, anchor=south east] at ({\cx+0.55*\dl},{\cy+0.55*\db})
           {$\rho$};
    
      % leaders to the pixel-margin labels
      \draw[guide] ({\xb+0.5*\dl},\ya) -- ({\xb+0.5*\dl},-0.58) -- (2.12,-0.58);
      \node[tny, text=ext, anchor=west] at (2.17,-0.58) {$n_{\lambda}$ px};
      \draw[guide] (\xa,{\yb+0.5*\db}) -- (-0.52,{\yb+0.5*\db}) -- (-0.75,1.84);
      \node[tny, text=ext, anchor=east] at (-0.80,1.84) {$n_{\beta}$ px};
    
      \node[font=\bfseries\small, anchor=north west] at (-3.75,-3.35)
           {(b) splitting it between the axes};
      \node[tny, anchor=north west, text width=7.5cm, align=left] at (-3.75,-3.90)
           {$\rho$ points radially outward from the disk centre (it is parallel
            to the region's own apparent position vector) and is the hypotenuse
            of a right triangle whose legs are the per-axis margins,
            $\Delta_{\lambda}=h|\sin\phi\cos\theta|$ and
            $\Delta_{\beta}=h|\sin\theta|$; dividing by the metres-per-pixel
            scale $m$ gives $n_{\lambda}$ and $n_{\beta}$.};
    \end{scope}
    
    % =====================================================================
    % Derivation strip
    % =====================================================================
    \node[anchor=north west, inner sep=7pt, draw=gray!45, line width=0.5pt, rounded corners=2pt, fill=gray!5, text width=16.2cm] at (-0.1,-6.35) {%
      \footnotesize
      In heliocentric Cartesian coordinates with $\hat{z}$ along the line of
      sight, $\hat{x}$ toward solar west and $\hat{y}$ toward solar north, the
      outward radial unit vector at longitude $\phi$ (measured from
      $\textsc{crln\_obs}$) and latitude $\theta$ is
      $\hat{r}=(\sin\phi\cos\theta,\allowbreak\ \sin\theta,\allowbreak\ \cos\phi\cos\theta)$.
      Raising the emission by $h$ along $\hat{r}$ ($h=2.5$~Mm at 1600 and
      1700\,\AA, $h=75$~Mm otherwise) displaces its apparent position by the
      tangential part of $h\hat{r}$, namely
      $\boldsymbol{\Delta}=(h\sin\phi\cos\theta,\allowbreak\ h\sin\theta)$, whence
      \begin{equation*}
        \underbrace{\psi=\arccos\!\big(\cos\phi\cos\theta\big)}
          _{\text{angle from disk centre}}
        \quad
        \underbrace{\rho=\|\boldsymbol{\Delta}\|= h\sin\psi}
          _{\text{apparent offset [m]}}
        \quad
        \underbrace{m=\textsc{cdelt1}\cdot
          \frac{\textsc{rsun\_ref}}{\textsc{rsun\_obs}}}
          _{\text{metres per pixel}}
        \quad
        \underbrace{n_{\lambda}=\Big\lceil
          \frac{h\,|\sin\phi\cos\theta|}{m}\Big\rceil,
          \ \,
          n_{\beta}=\Big\lceil\frac{h\,|\sin\theta|}{m}\Big\rceil}
          _{\text{pixel margin per axis }}
      \end{equation*}
      Both $\psi$ and $\rho$ cancel out of the per-axis expressions, so the margins follow from $\phi$, $\theta$ and $h$ alone and $\rho$ is left only as the consistency check $\sqrt{\Delta_{\lambda}^{2}+\Delta_{\beta}^{2}}=h\sin\psi$. Because $\boldsymbol{\Delta}$ is parallel to the region's apparent position vector, the offset is always radial in the image --- a quick check on any implementation.%
    };
    
    \end{tikzpicture}

    \caption{Computation of the pixel margin $(n_{\lambda},n_{\beta})$
      required to admit a coronal emission height $h$ above the
      photosphere. (a) The radially outward emission column at a region
      lying $\psi$ from the disk centre projects an apparent offset
      $\rho=h\sin\psi$ onto the plane of sky. (b) That offset is split
      between the image axes by its direction cosines, giving
      $\Delta_{\lambda}=h|\sin\phi\cos\theta|$ and
      $\Delta_{\beta}=h|\sin\theta|$, which are converted to pixels
      through the plate scale $m$ and added to the two bounding-box edges
      that face the nearest limb. }
    \label{fig:cutextend}
    
\end{figure}

%////////////////////Section\\\\\\\\\\\\\\\\\\\\\\\\\\\\\\
\section{Edge-detection classes in \texttt{edgedetection}}\classlabel{BaseEdgeDetector}\classlabel{CannyEdgeDetector}\classlabel{Gradient}\classlabel{GradientCalculator}

\texttt{BaseEdgeDetector} is the abstract base class for edge extraction over 2-D arrays. Its required method \texttt{get\_edges(source\_image)} is documented to return a binary edge/background array.\\

\noindent \texttt{CannyEdgeDetector} is the concrete edge detector documented on the public site. The class description explicitly lists the four implemented processing stages: Gaussian blur, Sobel-based horizontal and vertical edge detection, non-maximum suppression, and hysteresis thresholding. The constructor exposes \texttt{low\_threshold}, \texttt{high\_threshold}, and
\texttt{sigma}.\\

\noindent \texttt{Gradient} is the container class for gradient quantities, and \texttt{GradientCalculator} provides the documented methods \texttt{calculate\_gradient\_cart()},
\texttt{calculate\_gradient\_cart\_scipy()},\\
\texttt{calculate\_gradient\_polar()}, and \texttt{calculate\_gradient\_polar\_scipy()}. Its constructor selects among three documented operators: Prewitt, Sobel, and Roberts.
\\

%\noindent \texttt{PILDetector} provides \texttt{detect\_pil()} for polarity inversion line detection.

%==========================================================
\chapter{Image Texture and Statistical Parameters}\label{chap:imageparam}
%==========================================================

The library also provides image-parameter calculators that are adapted from the works of \cite{Ahmadzadeh_2019}, and that quantify statistical and texture-based properties of images. Such parameters are widely used in feature engineering because they convert complex pixel arrays into compact numerical descriptors.

%////////////////////Section\\\\\\\\\\\\\\\\\\\\\\\\\\\\\\
\section{Parameter Families}

The available descriptors include measures associated with:

\begin{itemize}
    \item Mean intensity
    \item Standard deviation
    \item Skewness
    \item Kurtosis
    \item Entropy
    \item Uniformity
    \item Relative smoothness
    \item Fractal-dimension-related behavior
    \item Texture contrast
    \item Directional texture properties
\end{itemize}

%////////////////////Section\\\\\\\\\\\\\\\\\\\\\\\\\\\\\\
\section{Importance in Space Weather Analysis}

These descriptors are valuable because solar images often exhibit nonuniform structure, fine-scale texture, and complex intensity distributions. Statistical and texture measures can help distinguish between quiet and active regions, summarize image complexity, and provide features for prediction or classification models.

%////////////////////Section\\\\\\\\\\\\\\\\\\\\\\\\\\\\\\
\section{Methodological Value}

From a methodological perspective, the inclusion of texture parameters reflects a recognized pattern in scientific image analysis: raw observations are transformed into robust, interpretable summary features that can be compared across instruments, time periods, and event classes.

%////////////////////Section\\\\\\\\\\\\\\\\\\\\\\\\\\\\\\
\section{Patch-based image-parameter classes in \texttt{imageparam}}\label{sec:imageparam_classes}

All calculators in \texttt{imageparam} share one interface. A calculator is constructed with a patch size and any parameter-specific settings, and \texttt{calculate\_parameter(data)} is then applied to a 2-D image such as an AIA cutout. The image is divided into non-overlapping square patches, the parameter is computed separately for each patch, and the result is a map with one value per patch. A $4096\times4096$ AIA image with $64\times64$-pixel patches gives a $64\times64$ parameter map, and calculating all ten parameters gives the $64\times64\times10$ data cube described in~\cite{Ahmadzadeh_2019}. With patch size \texttt{FULL}, the whole input is treated as a single patch and the result is one number. This is how each parameter is computed over a whole active-region cutout. Because cutouts contain NaN pixels outside the region mask (Section~\ref{sec:aia_cutout}), all of the calculators ignore NaN values.

%---------------------------------------------------------
\subsection{\texttt{BaseParamCalculator} and \texttt{PatchSize}}\classlabel{BaseParamCalculator}\classlabel{PatchSize}

\texttt{BaseParamCalculator} is the abstract parent of every image-parameter class. It implements \texttt{calculate\_parameter(data)}: it checks that the image width and height are both divisible by the patch size (raising \texttt{ValueError} otherwise), loops over the patches, and applies the subclass's per-patch function, which each subclass provides through its \texttt{calc\_func} property. A new parameter can therefore be added by writing only its per-patch function. \texttt{PatchSize} lists the allowed patch edge lengths in pixels: \texttt{ONE}, \texttt{FOUR}, \texttt{SIXTEEN}, \texttt{THIRTY\_TWO}, \texttt{SIXTY\_FOUR}, \texttt{ONE\_TWENTY\_EIGHT}, \texttt{TWO\_FIFTY\_SIX}, \texttt{FIVE\_TWELVE}, \texttt{TEN\_TWENTY\_FOUR}, and \texttt{FULL} for whole-image evaluation.

\medskip
\noindent In the definitions below, $z_j$ denotes the intensity of the $j$-th of the $n$ valid (non-NaN) pixels in a patch, and $\mu$ and $\sigma$ are their mean and standard deviation.

%---------------------------------------------------------
\subsection{\texttt{MeanParamCalculator}}\classlabel{MeanParamCalculator}
Computes the mean intensity of each patch, $\mu = \frac{1}{n}\sum_j z_j$. On a cutout this measures the overall brightness of the emission in the region, which in the EUV channels rises with the amount of heated coronal plasma.

%---------------------------------------------------------
\subsection{\texttt{StdDeviationParamCalculator}}\classlabel{StdDeviationParamCalculator}
Computes the standard deviation of intensity in each patch, $\sigma = \sqrt{\frac{1}{n}\sum_j (z_j-\mu)^2}$. It measures how much the brightness varies within the patch: it is high where bright loops and kernels sit next to dark background, and low in diffuse emission.

%---------------------------------------------------------
\subsection{\texttt{SkewnessParamCalculator}}\classlabel{SkewnessParamCalculator}
Computes the skewness of the intensity distribution in each patch,
\[ \gamma_1 = \frac{1}{n}\sum_j \frac{(z_j-\mu)^3}{\sigma^3}, \]
using \texttt{scipy.stats.skew}. Positive skewness means a long tail toward bright values, that is, a small number of pixels much brighter than the rest. Compact flare kernels and bright loop tops produce this signature.

%---------------------------------------------------------
\subsection{\texttt{KurtosisParamCalculator}}\classlabel{KurtosisParamCalculator}
Computes the excess (Fisher) kurtosis of each patch,
\[ \gamma_2 = \frac{1}{n}\sum_j \frac{(z_j-\mu)^4}{\sigma^4} - 3, \]
using \texttt{scipy.stats.kurtosis}. It measures how heavy the tails of the intensity distribution are compared with a Gaussian. Large values mean that most pixels are close to the mean and a few are extreme outliers.

%---------------------------------------------------------
\subsection{\texttt{RelativeSmoothnessParamCalculator}}\classlabel{RelativeSmoothnessParamCalculator}
Computes the relative smoothness $R = 1 - 1/(1+\sigma^2)$ of each patch. $R$ is 0 for a patch of constant intensity and approaches 1 as the variance grows, so it is a bounded, monotonic measure of contrast that does not depend on the units of the variance.

%---------------------------------------------------------
\subsection{\texttt{EntropyParamCalculator}}\classlabel{EntropyParamCalculator}
Computes the Shannon entropy of each patch's intensity histogram,
\[ E = -\sum_{i=1}^{N} p_i \log_2 p_i, \]
where $p_i$ is the fraction of the patch's pixels in bin $i$ of an $N$-bin histogram. The constructor sets the number of bins, \texttt{n\_bins} (or explicit bin edges), and the histogram range [\texttt{min\_val}, \texttt{max\_val}]. Pixels outside the range are ignored. The range is fixed rather than computed for each patch, so that entropies from different patches and images can be compared. Entropy measures how many intensity levels a patch uses: it is 0 for a uniform patch and $\log_2 N$ when all bins are equally populated. The histogram settings are hyper-parameters; values tuned for each AIA channel are given in the SWAN-SF-V2 dataset description.

%---------------------------------------------------------
\subsection{\texttt{UniformityParamCalculator}}\classlabel{UniformityParamCalculator}
Computes the uniformity (energy) of each patch's intensity histogram, $U = \sum_i p_i^2$. The histogram is built in the same way as for \texttt{EntropyParamCalculator}, with the same \texttt{n\_bins}, \texttt{min\_val}, and \texttt{max\_val} arguments. Uniformity is the complement of entropy: it is 1 when every pixel falls in the same bin and $1/N$ when all bins are equally populated.

%---------------------------------------------------------
\subsection{\texttt{TContrastParamCalculator}}\classlabel{TContrastParamCalculator}
Computes Tamura contrast~\cite{tamura1978textural} for each patch:
\[ C = \frac{\sigma^2}{\alpha_4^{1/4}}, \qquad \alpha_4 = \frac{\mu_4}{\sigma^4}, \]
where $\mu_4$ is the fourth central moment, so $\alpha_4$ is the (non-excess) kurtosis. Dividing the variance by the fourth root of the kurtosis lowers the contrast of patches whose large variance comes from a few outlying pixels. This makes $C$ follow the visual impression of contrast more closely than the variance does. A small constant is added to $\alpha_4$ to avoid division by zero, and patches whose contrast is undefined return 0.

%---------------------------------------------------------
\subsection{\texttt{FractalDimParamCalculator}}\classlabel{FractalDimParamCalculator}
Estimates the fractal (Minkowski--Bouligand, or box-counting) dimension of the edge structure in each patch. The constructor takes a patch size and an edge detector, normally a \texttt{CannyEdgeDetector} (Section~\ref{cls:CannyEdgeDetector}). For each patch:
(1) the edge detector converts the patch into a binary edge map;
(2) the patch is covered with square boxes of a sequence of sizes $s$, chosen from the patch size (e.g.\ $s=32,16,\dots,1$ for $64$- to $256$-pixel patches, and up to $s=64$ for larger patches and \texttt{FULL}), and the number of boxes $N(s)$ that contain at least one edge pixel is counted;
(3) a straight line is fitted by least squares to $\log N(s)$ against $\log s$, and the fractal dimension is the absolute value of its slope.
A smooth edge gives a dimension near 1, and edges that fill the plane increasingly give values approaching 2. The parameter therefore measures how complex and space-filling the structure in the patch is, such as the tangled loop systems of an evolving active region.

%---------------------------------------------------------
\subsection{\texttt{TDirectionalityParamCalculator} and \texttt{PeakDetector}}\classlabel{TDirectionalityParamCalculator}\classlabel{PeakDetector}
Computes Tamura directionality~\cite{tamura1978textural}, a measure of how strongly the intensity gradients in a patch are aligned along a few preferred directions. The constructor takes a \texttt{GradientCalculator} (Section~\ref{cls:GradientCalculator}), a \texttt{PeakDetector}, and a \texttt{quantization\_level}, the number of angular bins. For each patch:
(1) the gradient is computed in polar form, giving a magnitude and an angle at each pixel;
(2) gradient magnitudes below 15\% of the maximum in the patch are discarded, so that noise does not contribute, and the angles of the remaining pixels are binned into a normalized histogram over $[-\pi,\pi)$;
(3) the \texttt{PeakDetector} finds the dominant peaks of this histogram. It first finds every local maximum, then keeps peaks in order of height subject to a frequency threshold (absolute or percentile), a minimum distance between peaks, and an optional maximum number of peaks;
(4) each peak is assigned the angular range that extends half-way to its neighbouring peaks (wrapping around the circle), and the histogram-weighted sum of squared angular distances from each peak is accumulated over its range;
(5) this spread is divided by the spread of a uniform histogram with the same peaks, giving $F_{\mathrm{dir}} = 1 - \text{spread}/\text{spread}_{\mathrm{uniform}}$, clipped to $[0,1]$.
$F_{\mathrm{dir}}$ is 0 for an isotropic patch, or one with no significant peaks, and approaches 1 when all gradients point in a few sharply defined directions. In EUV images, high directionality picks out organized, aligned loop arcades and fan structures.

%==========================================================
\chapter{Magnetic Field Analysis}\label{chap:magfield}
%==========================================================

Magnetic structure is central to the physics of solar activity, as the energy stored in non-potential magnetic fields is the only source capable of powering strong solar transient events. \href{https://dmlab.cs.gsu.edu/docs/spaceweather_toolkit/}{\textbf{swdatatoolkit}} includes a dedicated magnetic field module that computes several important field-related quantities. This component is one of the library's principal scientific assets.

%////////////////////Section\\\\\\\\\\\\\\\\\\\\\\\\\\\\\\
\section{Available Magnetic Field Quantities}

The toolkit supports calculation of:

\begin{itemize}
    \item Horizontal magnetic field
    \item Vertical current-related field quantities
    \item Potential field extrapolations
    \item Total magnetic field
    \item Linear force-free field extrapolation
    \item Magnetic shear angle
\end{itemize}

%////////////////////Section\\\\\\\\\\\\\\\\\\\\\\\\\\\\\\
\section{Scientific Significance}

These calculations are relevant to studies of solar active regions, because magnetic complexity is closely associated with flare productivity and eruptive potential. The ability to derive field components and modeled fields from observational inputs allows researchers to investigate magnetic topology, non-potentiality, and field alignment.

%////////////////////Section\\\\\\\\\\\\\\\\\\\\\\\\\\\\\\
\section{Model-Based Field Computation}

The inclusion of potential and linear force-free field methods indicates that the library is not limited to direct measurement handling. It also supports physically motivated extrapolation and comparison between observed and idealized magnetic states. This is important for interpreting magnetic configurations in a solar context.

%////////////////////Section\\\\\\\\\\\\\\\\\\\\\\\\\\\\\\
\section{Magnetic field calculator classes in \texttt{magfield}}\label{sec:magfield_classes}

The \texttt{magfield} classes turn the SHARP vector magnetogram into the per-pixel field maps from which the scalar parameters of Chapter~\ref{chap:magparam} are computed. Their inputs follow the SHARP CEA conventions~\cite{bobra2014helioseismic, Hoeksema2014}. The data are remapped to a cylindrical equal-area grid, with rows indexing latitude and columns indexing longitude. The vertical (radial) field is $B_z = B_r$, the westward component is $B_x = B_\phi$, and the northward component is $B_y = -B_\theta$, where $B_\theta$ points south. Each component has a 1-$\sigma$ uncertainty map, all in Gauss. Inputs are \texttt{numpy} masked arrays. The caller is responsible for masking pixels that should not contribute, typically those outside the HARP bitmap, those with low disambiguation confidence, and NaNs. Outputs are masked arrays of the same shape, so the masks carry through to the parameter calculations. The data flow between these classes and the parameter calculators is summarized below.
\begin{center}
\small
\begin{tabular}{lll}
\toprule
Class & Physical inputs & Per-pixel outputs \\
\midrule
\texttt{HorizontalFieldCalculator} & $B_x, B_y$ (+ errors) & $B_h$, $\sigma_{B_h}$ \\
\texttt{TotalFieldCalculator} & $B_x, B_y, B_z$ (+ errors) & $|B|$, $\sigma_{|B|}$ \\
\texttt{VerticalCurrentFieldCalculator} & $B_x, B_y$ (+ errors) & $J_z$, $\sigma_{J_z}$, $\partial B_y/\partial x$, $\partial B_x/\partial y$ \\
\texttt{GreenpotCalculator} & $B_z$ & $B_x^{p}, B_y^{p}$ (JSOC SHARP replica) \\
\texttt{GreenFunctionPotentialCalculator} & $B_z$, window & $B_x^{p}, B_y^{p}$ (potential field) \\
\texttt{LinearForceFreeFieldCalculator} & $B_z$, $\alpha$ & $B_x^{p}, B_y^{p}$ (LFFF; potential for $\alpha=0$) \\
\texttt{ShearAngleFieldCalculator} & $B_x, B_y, B_z, J_z, B_x^{p}, B_y^{p}$ & signed $\sin\Theta$ \\
\bottomrule
\end{tabular}
\end{center}

%---------------------------------------------------------
\subsection{\texttt{HorizontalFieldCalculator}}\classlabel{HorizontalFieldCalculator}
Computes the horizontal field strength $B_h = \sqrt{B_x^2 + B_y^2}$ at each pixel, together with its uncertainty, propagated in quadrature from the errors of $B_x$ and $B_y$. $B_h$ measures how much of the photospheric field lies along the surface rather than pointing up or down. It is an input to the mean inclination angle (\texttt{GammaCalculator}) and the horizontal-gradient parameter (\texttt{HorizontalDerivativeCalculator}).

%---------------------------------------------------------
\subsection{\texttt{TotalFieldCalculator}}\classlabel{TotalFieldCalculator}
Computes the total field strength $|B| = \sqrt{B_x^2 + B_y^2 + B_z^2}$ at each pixel and its propagated 1-$\sigma$ uncertainty. The total-field map is the input to \texttt{TotalDerivativeCalculator} (\textsc{meangbt}).

%---------------------------------------------------------
\subsection{\texttt{VerticalCurrentFieldCalculator}}\classlabel{VerticalCurrentFieldCalculator}
Computes the vertical current density from Amp\`ere's law, $J_z \propto (\nabla\times\mathbf{B})_z = \partial B_y/\partial x - \partial B_x/\partial y$. The derivatives are evaluated with central differences on the CEA grid, which is equivalent to integrating $\mathbf{B}$ around the boundary of each pixel and dividing by the pixel area. Following the SHARP convention, $J_z$ is left in Gauss per pixel; the conversion to A\,m$^{-2}$ (a factor $1/\mu_0$ and the pixel size in meters) is applied by the parameter calculators that report physical currents. The method returns the $J_z$ map, its 1-$\sigma$ uncertainty (propagated from the errors of $B_x$ and $B_y$), and the two intermediate derivative maps. The outermost row and column of the array, and any pixel whose stencil reaches a masked or NaN neighbour, are masked in the output, because the derivative cannot be evaluated there. The masks of the input arrays should therefore mark only invalid values, as stated in the docstring: the bitmap and disambiguation masks are applied afterwards to the result, so that pixels just inside the region boundary are not lost. $J_z$ maps are used in the twist ($\alpha$), current-helicity, total-current, and polarity-current parameters, and to sign the shear (below).

%---------------------------------------------------------
\subsection{\texttt{GreenpotCalculator}}\classlabel{GreenpotCalculator}
Reproduces the potential field of the JSOC SHARP pipeline (the \texttt{greenpot} routine of \texttt{sw\_functions.c}, also in the Python port of \cite{bobra2014helioseismic}), so that the toolkit's potential-field parameters can be compared with the SHARP header keywords. The \emph{potential} (current-free) field $\mathbf{B}^p = -\nabla\Phi$ is the minimum-energy field with the observed vertical component at the photosphere, and is the reference against which non-potentiality is measured. The JSOC code computes the scalar potential as
\[ \Phi(x,y) = \sum_{x',y'} B_z(x',y')\,\frac{\delta z}{\sqrt{(x-x')^2+(y-y')^2+\delta z^2}}, \]
with a ``monopole depth'' $\delta z=10^{-3}$\,pixel, summed over a $21\times21$-pixel window ($\pm10$ pixels). The horizontal field is then $B_x^p = -\partial\Phi/\partial x$ and $B_y^p = -\partial\Phi/\partial y$, evaluated with central differences. The implementation uses a vectorized sliding-window product. The JSOC loop bounds make the window asymmetric ($-10$ to $+9$ pixels), whereas the window here is centred on the pixel; with the same inputs, the two give \textsc{meanshr} values that differ by about 0.03\%. NaN pixels in $B_z$ are treated as zero flux, the potential is set to zero there, and they are masked in the output. Maps narrower than 20 pixels in either direction return zero fields.

The JSOC source describes this as ``a simplified definition'', and it is not an accurate solution of the potential-field problem. The Green's-function solution (Section~\ref{cls:GreenFunctionPotentialCalculator}) weights a neighbour at distance $r$ by $1/r$ and the pixel's own flux by the integral of $1/r$ over the pixel, about 3.5. The kernel above instead samples $1/\sqrt{r^2+\delta z^2}$ at the pixel centre, which gives the pixel's own flux a weight of $1/\delta z=1000$. Relative to its neighbours, the central pixel is therefore overweighted by a factor of about 280, and the other 440 pixels of the window together carry only about 7\% of its weight. The potential is then close to $B_z$ itself (in G rather than G\,pixel), and the horizontal field is close to $-\nabla B_z$ evaluated pixel to pixel, which is much weaker than the true potential field (Section~\ref{sec:potential_model_comparison}). The factor $\delta z$ only rescales the sum and does not change this balance. The truncation to $\pm10$ pixels removes most of the contribution of distant flux as well. The SHARP keywords that use this field (\textsc{meanshr}, \textsc{shrgt45}, and the free-energy keywords \textsc{meanpot} and \textsc{totpot}) inherit its weakness. The class is kept for compatibility with those keywords. For a physical potential field, use \texttt{GreenFunctionPotentialCalculator} or \texttt{LinearForceFreeFieldCalculator} with $\alpha=0$.

%---------------------------------------------------------
\subsection{\texttt{GreenFunctionPotentialCalculator}}\classlabel{GreenFunctionPotentialCalculator}
Computes the potential field at the boundary $z=0$ from the Green's-function solution of the Neumann problem for the half-space $z\ge0$~\cite{sakurai1982green}:
\[ \Phi(x,y) = \frac{1}{2\pi}\sum_{x',y'} B_z(x',y')\,G(x-x',y-y'), \qquad G(r)=\begin{cases} 1/r, & r>0,\\ 4\ln\!\left(1+\sqrt2\right)\approx3.53, & r=0,\end{cases} \]
with distances in pixels. $G(0)$ is the integral of $1/r$ over a unit pixel centred on the origin, so the pixel's own flux is weighted as a uniformly magnetized pixel rather than a point source. As before, $B_x^p=-\partial\Phi/\partial x$ and $B_y^p=-\partial\Phi/\partial y$ are central differences (one-sided at the edges). The sum is evaluated by FFT convolution with zero padding (\texttt{scipy.signal.fftconvolve}), so the boundary is open: flux outside the map is taken to be zero, and there are no periodic images. NaN pixels in $B_z$ are treated as zero flux and masked in the output.

The constructor argument \texttt{window} sets the half-width of a square kernel ($-\texttt{window}$ to $+\texttt{window}$ pixels in each direction). The default, \texttt{None}, uses the whole map, so that every pixel contributes to every other one. Because $G$ falls off only as $1/r$, a truncated kernel loses much of the field of distant flux. Section~\ref{sec:potential_model_comparison} shows the effect of the window on a real region: with the full map the result agrees closely with the Fourier potential field, and with $\pm10$ pixels, as used by the SHARP pipeline, it does not. The unit tests check the class against a brute-force evaluation of the sum, and against the analytic boundary field of a pair of opposite monopoles buried below the surface, which it reproduces to within 5\% in RMS strength and 3\textdegree\ in direction with the full-map kernel.

%---------------------------------------------------------
\subsection{\texttt{LinearForceFreeFieldCalculator}}\classlabel{LinearForceFreeFieldCalculator}
Computes the horizontal boundary field of a \emph{linear force-free} field, $\nabla\times\mathbf{B} = \alpha\mathbf{B}$ with constant $\alpha$, in the half-space $z\ge 0$ that has the observed $B_z$ at $z=0$ and decays with height ($\hat{\mathbf B}\propto e^{-qz}$). The calculation is done in Fourier space. With wavenumbers $(k_x,k_y)$, $k^2=k_x^2+k_y^2$, $q=\sqrt{k^2-\alpha^2}$, and the transform convention $f(x)=\sum_k \hat f(k)\,e^{+ikx}$ used by \texttt{numpy.fft},
\[ \hat B_x = -i\,\frac{k_x q - \alpha k_y}{k^2}\,\hat B_z, \qquad \hat B_y = -i\,\frac{k_y q + \alpha k_x}{k^2}\,\hat B_z . \]
These satisfy $\nabla\cdot\mathbf B = ik_x\hat B_x + ik_y\hat B_y - q\hat B_z = 0$ and $(\nabla\times\mathbf B)_z = \alpha B_z$.
The $k=0$ (net flux) mode is set to zero, and for $k^2<\alpha^2$ the complex $q$ is kept and the real part of the inverse FFT is returned. The constructor takes \texttt{alpha} (in inverse units of the pixel spacing \texttt{dx}, \texttt{dy}; with unit spacing, 1/pixel) and the pixel spacings. With the default $\alpha=0$ the relations reduce to the potential field $\mathbf B=-\nabla\Phi$, $\hat B_{x,y} = -i k_{x,y}\hat B_z/k$, whose horizontal component points away from positive and towards negative flux, so the class can be used as a potential-field model. Like the full-map \texttt{GreenFunctionPotentialCalculator}, the Fourier solution is global: every pixel's potential field depends on the whole patch. Unlike it, the patch is implicitly treated as periodic, so the solution is sensitive to flux imbalance and to strong fields near the patch edges. NaNs are treated as zero flux and masked in the output, as in the other potential-field classes. For the region in Section~\ref{sec:potential_model_comparison}, the Fourier and full-map Green's-function fields agree closely, and both differ strongly from \texttt{GreenpotCalculator}.

%---------------------------------------------------------
\subsection{\texttt{ShearAngleFieldCalculator}}\classlabel{ShearAngleFieldCalculator}
Produces a per-pixel map of magnetic shear, that is, how far the observed field direction departs from the potential field at each location. The potential field is assumed to share the observed vertical component ($B_z^p = B_z$), and the 3-D shear angle is
\[ \Theta = \arccos\!\left(\frac{B_x^p B_x + B_y^p B_y + B_z^2}{|\mathbf{B}^p|\,|\mathbf{B}|}\right). \]
The map returned is not $\Theta$ itself but
\[ S = \operatorname{sign}(J_z/B_z)\,\sin\Theta \quad\text{where } |B_z|\ge B_{\mathrm{thr}},\qquad S=0 \text{ otherwise.} \]
$\sin\Theta$ is 0 where the field is potential and 1 where it is sheared by $90^\circ$. The sign of $J_z/B_z$ (the sign of the local force-free twist $\alpha$) separates right-handed from left-handed twist. The threshold $B_{\mathrm{thr}}$ (constructor argument \texttt{bz\_threshold}, default 100\,G) suppresses weak-field pixels, where the field direction is dominated by noise. The inputs are the observed $B_x, B_y, B_z$, the $J_z$ map from \texttt{VerticalCurrentFieldCalculator}, and $B_x^p, B_y^p$ from either potential-field model.

The signed map can be used directly, for example by \texttt{HelicityGuessCalculator} (Section~\ref{cls:HelicityGuessCalculator}). Its magnitude $w = |S| \in [0,1]$ can be used as a per-pixel \emph{weight matrix}: passed as the optional \texttt{weight\_matrix} argument to the parameter calculators of Chapter~\ref{chap:magparam}, it gives shear-weighted versions of the SHARP parameters, in which strongly sheared, non-potential parts of the region count fully and potential parts count little or not at all (Section~\ref{sec:shear_weighted}).

%////////////////////Section\\\\\\\\\\\\\\\\\\\\\\\\\\\\\\
\section{Choice of potential-field model}\label{sec:potential_model_comparison}

{\sloppy The toolkit has three ways of computing the potential field from the observed $B_z$. \texttt{GreenpotCalculator} reproduces the JSOC SHARP routine. The class \texttt{GreenFunctionPotentialCalculator} is a Green's-function solution with a configurable window. The third is \texttt{LinearForceFreeFieldCalculator} with $\alpha=0$, a Fourier solution on a periodic patch. All three keep the observed $B_z$ at the boundary, but their horizontal fields differ, and so does any quantity defined relative to the potential state. This section compares them for HARP~11149 (NOAA AR~13664) at 2024-05-11 01:24:00~TAI, about one minute after the peak of the X5.8 flare produced by this region. All statistics are over the 137{,}591 pixels with $|B_z|\ge100$\,G that pass the SHARP quality mask.\par}

\begin{figure}[p]
    \centering
    \includegraphics[width=\textwidth]{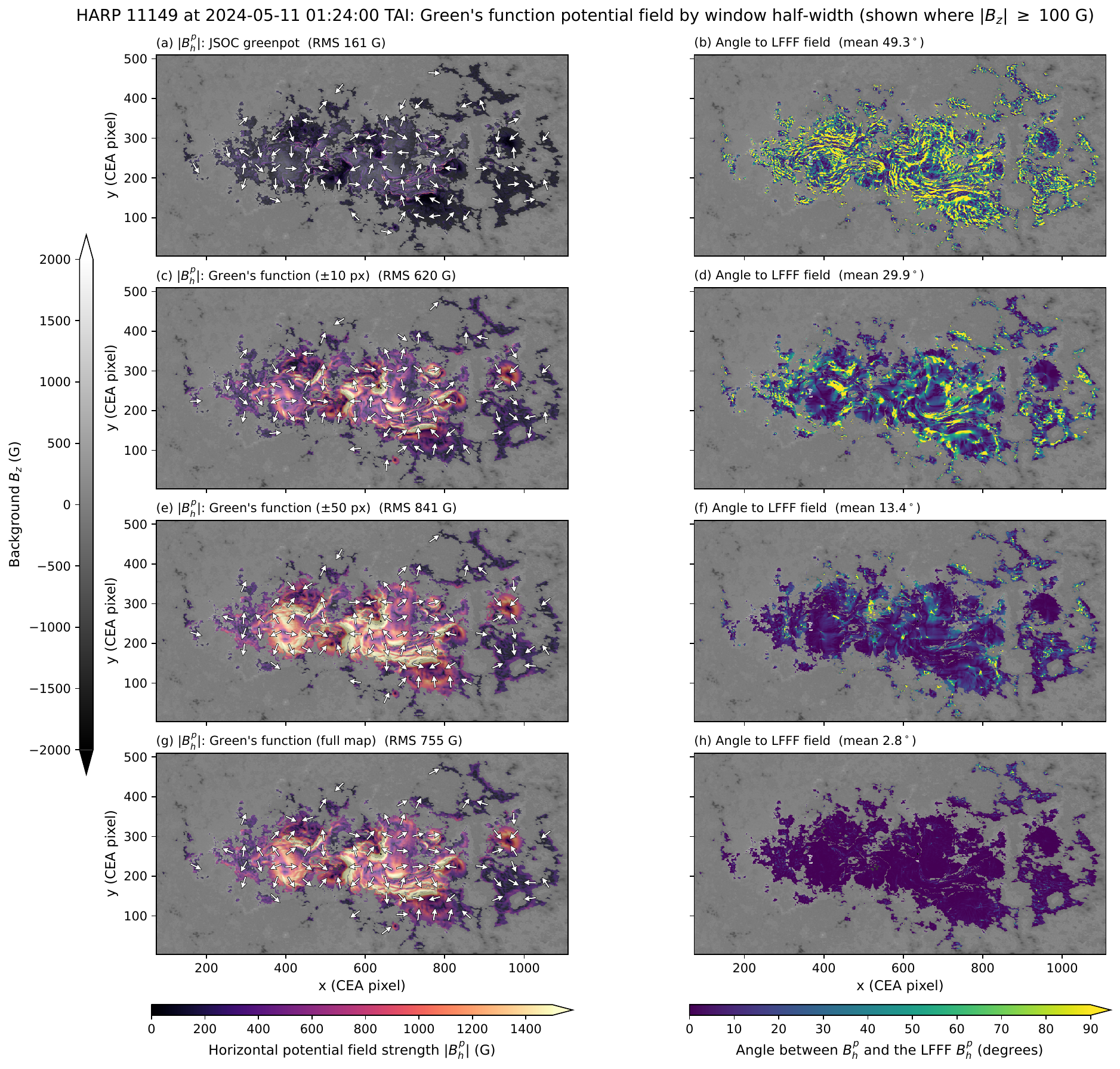}
    \caption{Horizontal potential field of HARP~11149 at 2024-05-11 01:24:00~TAI from (a) \texttt{GreenpotCalculator} and from \texttt{GreenFunctionPotentialCalculator} with a window of (c) $\pm10$ pixels, (e) $\pm50$ pixels, and (g) the full map. The left panels show the strength $|B_h^p|$, with arrows of fixed length every 40 pixels showing its direction. The right panels (b, d, f, h) show the angle between each field and the Fourier potential field from \texttt{LinearForceFreeFieldCalculator} ($\alpha=0$), whose strength is shown in Figure~\ref{fig:potential_compare}b. The overlays are drawn where $|B_z|\ge100$\,G and the pixel passes the SHARP quality mask, on a gray background of $B_z$ ($\pm2000$\,G). Axes are CEA pixels (0.03\textdegree, about 0.36\,Mm).}
    \label{fig:green_windows}
\end{figure}

\paragraph{Green's-function window (Figure~\ref{fig:green_windows}, Table~\ref{tab:green_windows}).} Figure~\ref{fig:green_windows} shows the horizontal potential field of \texttt{GreenpotCalculator} and of \texttt{GreenFunctionPotentialCalculator} with windows of $\pm10$ pixels, $\pm50$ pixels, and the full map, together with the angle between each field and the Fourier field. The \texttt{GreenpotCalculator} field is weak almost everywhere (panel a), with an RMS of 161\,G, against 766\,G for the Fourier field and 699\,G for the observed horizontal field. It is strong only along thin lines at the umbral boundaries and the central polarity inversion line, where $B_z$ changes most steeply. This is the result of the kernel described in Section~\ref{cls:GreenpotCalculator}: because the central pixel carries most of the weight, $\Phi\approx B_z$, and the horizontal field is close to $-\nabla B_z$. For this record $|B_h^p|$ from \texttt{GreenpotCalculator} and $|\nabla B_z|$ (in G per pixel) have a pixel-wise correlation of 0.9998, and their directions differ by 1.1\textdegree\ on average. Its direction is also far from that of the Fourier field (mean angle 49\textdegree, panel b).

With the correct kernel, the field is much stronger even with the same $\pm10$-pixel window (RMS 620\,G, panel c), and its direction is closer to that of the Fourier field (mean angle 30\textdegree, panel d). It is still too weak in and around the large spots, where much of the field comes from flux more than 10 pixels (3.6\,Mm) away. Its median strength is 0.75 of the Fourier value. With $\pm50$ pixels the strength is about right on average (median ratio 1.09), and the mean angle falls to 13\textdegree. With the full map the Green's-function and Fourier fields are nearly the same (panels g and h): the pixel-wise correlation of their strengths is 0.998, the median ratio is 0.99, and their directions differ by 2.8\textdegree\ on average. The remaining differences come from the boundary: the Green's-function solution has an open boundary with no flux outside the map, and the Fourier solution a periodic one. Two different solutions of the same problem therefore agree once the kernel and window are correct. The poor agreement of \texttt{GreenpotCalculator} with the Fourier field is a property of the JSOC implementation, not of the Green's-function method.

Compared with the observed horizontal field, the full-map Green's-function field and the Fourier field also score the same (strength correlation 0.77, mean angle 46\textdegree). The $\pm10$-pixel field is slightly worse (0.70 and 51\textdegree), and the \texttt{GreenpotCalculator} field much worse (0.33 and 61\textdegree). The observed field is of course not potential, so these numbers are a check of plausibility, not of accuracy.

\begin{table}[htb!]
\centering
\small
\caption{Horizontal potential field of HARP~11149 at 2024-05-11 01:24:00~TAI by model, over the 137{,}591 pixels with $|B_z|\ge100$\,G. ``Corr.'' is the pixel-wise correlation of $|B_h^p|$ with the Fourier or the observed $|B_h|$; ``Ratio'' is the median of $|B_h^p|$ divided by the Fourier value; ``Angle'' is the mean angle between the horizontal field vectors. The RMS of the Fourier field is 766\,G, and of the observed field 699\,G.}
\label{tab:green_windows}
\begin{tabular}{lrrrrrr}
\toprule
 & & \multicolumn{3}{c}{vs.\ Fourier ($\alpha=0$)} & \multicolumn{2}{c}{vs.\ observed} \\
\cmidrule(lr){3-5}\cmidrule(lr){6-7}
Model & RMS (G) & Corr. & Ratio & Angle & Corr. & Angle \\
\midrule
\texttt{GreenpotCalculator} (JSOC) & 161 & 0.49 & 0.18 & 49.3\textdegree & 0.33 & 61.1\textdegree \\
Green's function, $\pm10$ px & 620 & 0.85 & 0.75 & 29.9\textdegree & 0.70 & 51.0\textdegree \\
Green's function, $\pm50$ px & 841 & 0.96 & 1.09 & 13.4\textdegree & 0.76 & 46.9\textdegree \\
Green's function, full map & 755 & 0.998 & 0.99 & 2.8\textdegree & 0.77 & 46.5\textdegree \\
\bottomrule
\end{tabular}
\end{table}

\begin{figure}[p]
    \centering
    \includegraphics[height=0.78\textheight]{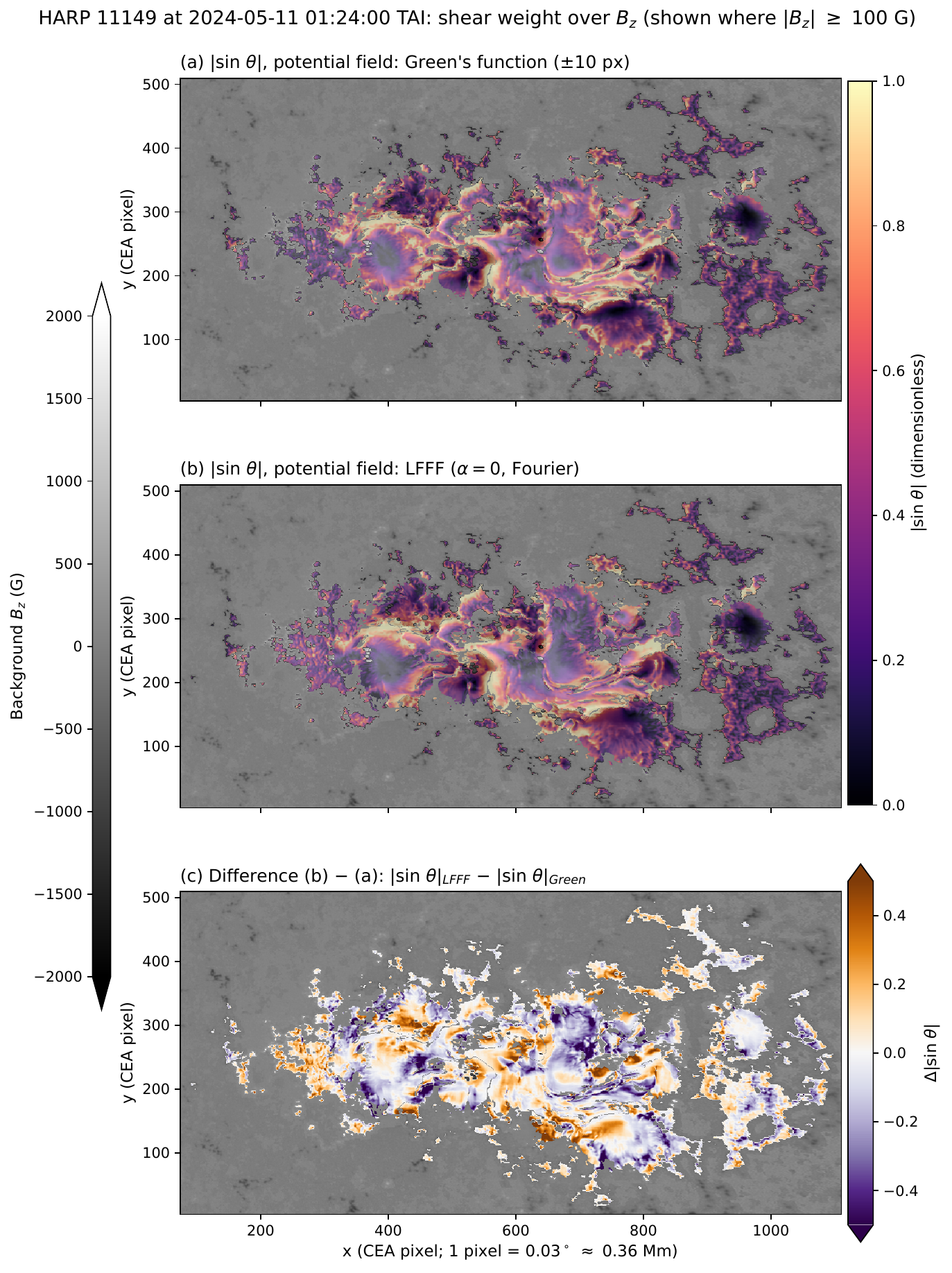}
    \caption{Shear magnitude $|\sin\Theta|$ for HARP~11149 at 2024-05-11 01:24:00~TAI, computed by \texttt{ShearAngleFieldCalculator} with the potential field from (a) \texttt{GreenFunctionPotentialCalculator} with a $\pm10$-pixel window and (b) \texttt{LinearForceFreeFieldCalculator} ($\alpha=0$). Panel (c) shows the difference (b)$-$(a). The overlays are drawn where $|B_z|\ge100$\,G and the pixel passes the SHARP quality mask (\textsc{conf\_disambig}~$\ge70$, bitmap~$\ge30$), on a gray background of $B_z$ ($\pm2000$\,G). Axes are CEA pixels (0.03\textdegree, about 0.36\,Mm).}
    \label{fig:shear_compare}
\end{figure}

\begin{figure}[p]
    \centering
    \includegraphics[height=0.85\textheight]{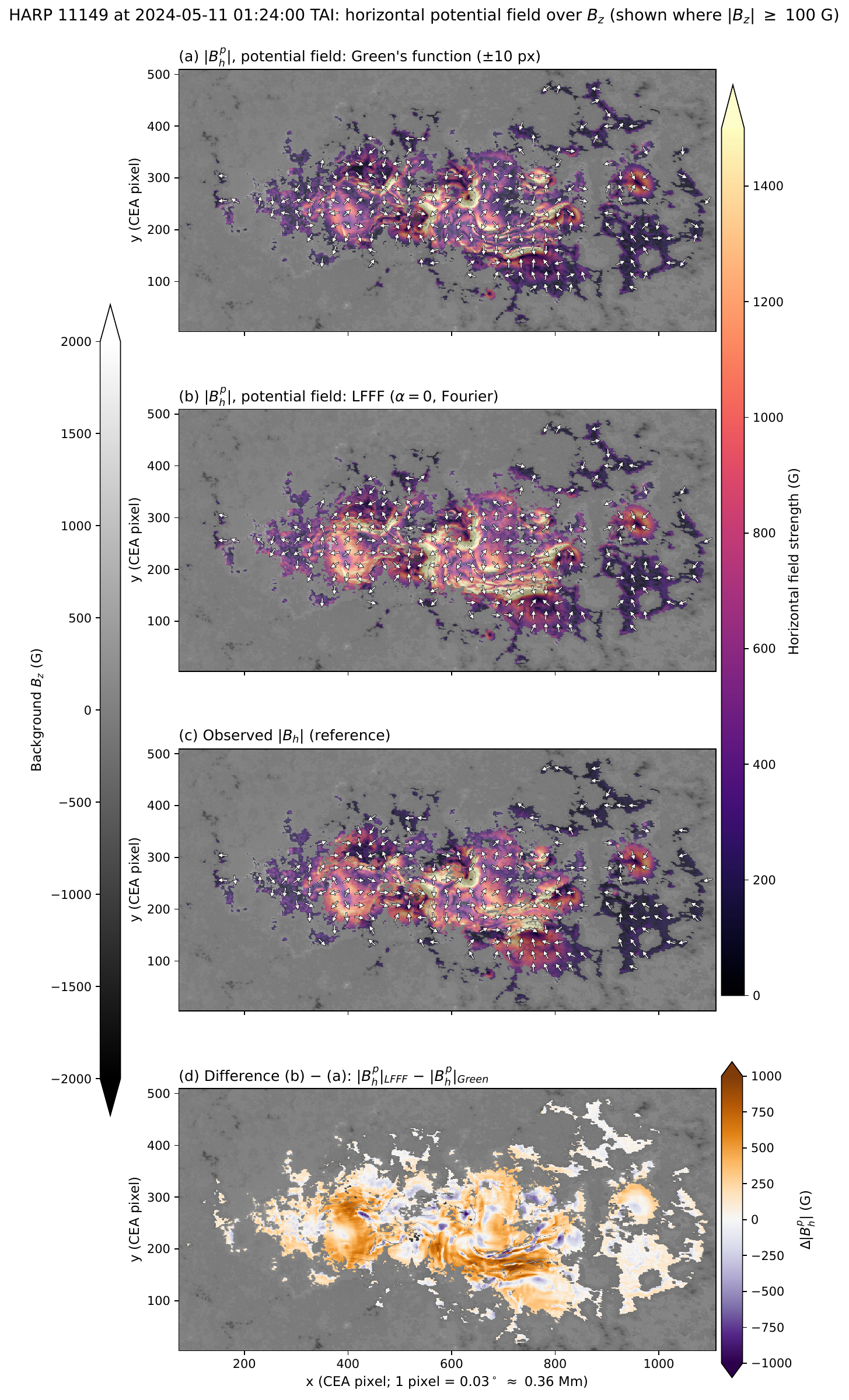}
    \caption{Horizontal potential field strength $|B_h^p|=\sqrt{(B_x^p)^2+(B_y^p)^2}$ for HARP~11149 at 2024-05-11 01:24:00~TAI, from (a) \texttt{GreenFunctionPotentialCalculator} with a $\pm10$-pixel window and (b) \texttt{LinearForceFreeFieldCalculator} ($\alpha=0$), with (c) the observed horizontal field strength $|B_h|$ for reference. Panel (d) shows the difference (b)$-$(a). Arrows show the direction of the horizontal field every 24 pixels; their length is fixed, and the strength is given by the color. The pixels shown, the background, and the axes are as in Figure~\ref{fig:shear_compare}.}
    \label{fig:potential_compare}
\end{figure}

\paragraph{Green's function with $\pm10$ pixels versus Fourier (Figures~\ref{fig:shear_compare} and~\ref{fig:potential_compare}).} The SWAN-SF-V2 pipeline (\texttt{aia-mvts}) computes every potential-field-dependent parameter twice: with \texttt{GreenFunctionPotentialCalculator} using the $\pm10$-pixel window of the SHARP pipeline, and with the Fourier model. Figure~\ref{fig:potential_compare} compares the two potential fields. The Fourier field is the stronger of the two in 75\% of the pixels (panel d), by 143\,G on average (medians 370\,G and 521\,G, against 437\,G for the observed field). The differences are largest in and around the big spots, where the $\pm10$-pixel window misses the field of distant flux. Elsewhere, and in the weak-field plage, the two fields are similar. The pixel-wise correlation of the strengths is 0.85, and the mean angle between the fields is 30\textdegree.

Figure~\ref{fig:shear_compare} shows the resulting shear magnitude $|S|=|\sin\Theta|$ from \texttt{ShearAngleFieldCalculator}. Both maps show strongly sheared field along the central polarity inversion line and around the penumbrae, and have a similar mean (0.41 with the Green's-function and 0.39 with the Fourier potential field). The pixel-wise correlation between the maps is 0.77 and the mean absolute difference is 0.13. The differences are of both signs and are mostly in and around the spots (panel c), where the two potential fields have different directions. With \texttt{GreenpotCalculator}, by contrast, the mean shear is 0.51, because its nearly vertical potential field makes any inclined field, such as a penumbra, appear sheared, and the correlation with the Fourier shear map is only 0.57.

%==========================================================
\chapter{Magnetic-Parameter Derivation}\label{chap:magparam}
%==========================================================

In addition to magnetic-field components, the toolkit includes a substantial set of higher-level magnetic parameters. These descriptors are typically derived from vector or line-of-sight magnetic data and are frequently used as inputs to statistical or machine-learning models.

%////////////////////Section\\\\\\\\\\\\\\\\\\\\\\\\\\\\\\
\section{Types of Derived Parameters}

The available parameter families include quantities related to:

\begin{itemize}
    \item Magnetic flux
    \item Force-free alpha ($\alpha$)
    \item Magnetic shear
    \item Helicity
    \item Lorentz force
    \item Polarity structure
    \item Gradient and derivative measures
    \item Free magnetic energy
    \item Vertical current
    \item $R$-value and shear-weighted $R$-value
    \item Line-of-sight, horizontal, total, and vertical derivatives
    \item Effective connected magnetic field ($B_\text{eff}$) from a magnetic charge topology model
\end{itemize}

%////////////////////Section\\\\\\\\\\\\\\\\\\\\\\\\\\\\\\
\section{Analytical Relevance}

These metrics are important because they summarize the physical complexity of an active region in forms that are easier to compare and model than full-field maps. They are widely used in flare forecasting, active-region classification, and studies of magnetic non-potentiality.

% ----------------------------------------------------------------
\section{Magnetic Field Parameter Calculators (\texttt{magparam})}
\label{sec:magparam}
% ----------------------------------------------------------------

The \texttt{magparam} module provides NumPy implementations of the SHARP active-region parameters~\cite{bobra2014helioseismic}, adapted from Bobra's Python reference implementation of the JSOC SHARP module, together with several parameters that are not in the SHARP series. Each calculator reduces the per-pixel field maps of Chapter~\ref{chap:magfield} (or the line-of-sight magnetogram) to a small set of scalars describing the whole region. Each scalar comes with a propagated 1-$\sigma$ uncertainty, computed by adding per-pixel variances on the assumption that pixel errors are independent. The \texttt{magparam.mct} sub-package, which computes the effective connected magnetic field $B_\text{eff}$ from a model of the region's flux connectivity rather than from per-pixel sums, is described separately in Section~\ref{sec:mct}.

\paragraph{Common inputs.} Unless stated otherwise, the calculators take:
\begin{itemize}[nosep]
    \item \emph{field maps}: the CEA field components $B_x$, $B_y$, $B_z$ in Gauss and their 1-$\sigma$ error maps, or maps derived from them ($B_h$, $|B|$, $J_z$, $B^p_{x,y}$), as masked arrays in which pixels outside the region of interest are masked;
    \item \emph{plate-scale information} from the FITS header, used to convert per-pixel sums into physical units: the reference solar radius \textsc{rsun\_ref} (m), the apparent solar radius \textsc{rsun\_obs} (arcsec), and the plate scale \textsc{cdelt1} in arcsec\,pixel$^{-1}$ (for the CEA series this must be converted from degrees, as shown in the \texttt{FluxCalculator} docstring). Together these give the pixel size $\delta = \textsc{cdelt1}\cdot\textsc{rsun\_ref}/\textsc{rsun\_obs}$ in meters and the pixel area $dA=\delta^2$;
    \item an optional \emph{weight matrix} $w$ of the same shape as the field maps, which multiplies each pixel's contribution to the sums (and to the variance sums). With no weight matrix every valid pixel has weight 1, which reproduces the standard SHARP keywords. Passing the shear magnitude from \texttt{ShearAngleFieldCalculator} gives the shear-weighted variants discussed in Section~\ref{sec:shear_weighted}.
\end{itemize}
\paragraph{Common output.} Each \texttt{calc} method returns a single-row \texttt{pandas} \texttt{DataFrame}, one column per parameter and per uncertainty, named after the SHARP keywords so that results can be concatenated over time into a multivariate time series. If no valid pixels remain, the parameters are reported as zero (or NaN where stated) rather than raising an exception.

\medskip
\noindent In the formulas below, sums run over the valid (unmasked) pixels $i$ and $N$ is their number; the per-pixel weights $w_i$ are omitted for clarity.

%---------------------------------------------------------
\subsection{\texttt{FluxCalculator}}\classlabel{FluxCalculator}
Computes the total unsigned magnetic flux and the net (signed) flux of the region:
\[ \textsc{usflux} = \sum_i |B_{z,i}|\,dA, \qquad \textsc{fluximb} = \sum_i B_{z,i}\,dA, \]
in Maxwell (G\,cm$^2$). The uncertainty \textsc{errvf} is $\sqrt{\sum_i\sigma_{B_z,i}^2}\,dA$. \textsc{usflux} measures the size and total field strength of the region and is one of the parameters most strongly correlated with flare productivity. \textsc{fluximb} measures how unbalanced the two polarities are. \texttt{calc()} uses the radial field $B_z$, which is already corrected for projection because the data are on the CEA grid. \texttt{calc\_los()} applies the same sums to a line-of-sight magnetogram and returns no uncertainty.

%---------------------------------------------------------
\subsection{\texttt{GammaCalculator}}\classlabel{GammaCalculator}
Computes the mean inclination of the field relative to the vertical,
\[ \textsc{meangam} = \frac{1}{N}\sum_i \arctan\frac{B_{h,i}}{|B_{z,i}|} \]
in degrees, and its uncertainty \textsc{errgam}. The inputs are the $B_z$ and $B_h$ maps and their errors (from \texttt{HorizontalFieldCalculator}). Values near $90^\circ$ mean the field lies mostly along the surface, which is typical of sheared, flux-emerging regions. Values near $0^\circ$ mean the field is mostly vertical, as in sunspot umbrae.

%---------------------------------------------------------
\subsection{Field-gradient calculators}\classlabel{BaseDerivativeCalculator}\classlabel{TotalDerivativeCalculator}\classlabel{HorizontalDerivativeCalculator}\classlabel{ZDerivativeCalculator}\classlabel{LOSDerivativeCalculator}
\texttt{BaseDerivativeCalculator} implements the shared algorithm for the SHARP mean-gradient parameters, and four subclasses apply it to different field maps. For a field map $B$, the local gradient magnitude $|\nabla B| = \sqrt{(\partial B/\partial x)^2 + (\partial B/\partial y)^2}$ is computed with central differences at every interior pixel whose neighbours are all valid. Its mean is then converted from G\,pixel$^{-1}$ to G\,Mm$^{-1}$. Edge pixels, and pixels next to masked or NaN pixels, are excluded.
\begin{itemize}[nosep]
    \item \texttt{TotalDerivativeCalculator}: input $|B|$ (from \texttt{TotalFieldCalculator}), output \textsc{meangbt}/\textsc{errbt}.
    \item \texttt{HorizontalDerivativeCalculator}: input $B_h$, output \textsc{meangbh}/\textsc{errbh}.
    \item \texttt{ZDerivativeCalculator}: input $B_z$, output \textsc{meangbz}/\textsc{errbz}.
    \item \texttt{LOSDerivativeCalculator}: input the line-of-sight field $B_{\mathrm{los}}$, output \textsc{meangbl}/\textsc{errbl}.
\end{itemize}
Steep field gradients occur where strong opposite polarities are pressed together. Such compact, high-gradient polarity inversion lines are a well-known signature of flare-productive regions.

%---------------------------------------------------------
\subsection{\texttt{TotalVerticalCurrentCalculator}}\classlabel{TotalVerticalCurrentCalculator}
Computes the total unsigned vertical current and the mean vertical current density from the $J_z$ map of \texttt{VerticalCurrentFieldCalculator}:
\[ \textsc{totusjz} = \sum_i |J_{z,i}|\,dA \;\;[\mathrm{A}], \qquad \textsc{meanjzd} = \frac{1}{N}\sum_i J_{z,i} \;\;[\mathrm{mA\,m^{-2}}], \]
with uncertainties \textsc{errusi} and \textsc{errjz}. The per-pixel $J_z$, which is in Gauss per pixel, is converted to physical units using $1/\mu_0$ and the pixel size. \textsc{totusjz} measures the total amount of electric current flowing through the photosphere in the region, and hence the departure of the field from a potential (current-free) state. Together with \textsc{usflux}, it is among the strongest single predictors in SHARP-based flare forecasting.

%---------------------------------------------------------
\subsection{\texttt{AlphaCalculator}}\classlabel{AlphaCalculator}
Computes the best-fit single value of the force-free twist parameter, using the $B_z^2$-weighted estimator,
\[ \textsc{meanalp} = \frac{\sum_i J_{z,i}\,B_{z,i}}{\sum_i B_{z,i}^2}\times C \;\;[\mathrm{Mm^{-1}}], \]
where $C$ converts from pixel$^{-1}$ to Mm$^{-1}$. The uncertainty is \textsc{erralp}. In a force-free field $\nabla\times\mathbf{B}=\alpha\mathbf{B}$, and $\alpha$ measures how tightly the field lines are twisted. Its sign gives the handedness of the twist. Weighting by $B_z^2$ means that strong-field pixels, where $J_z/B_z$ is well determined, dominate the estimate. The result is NaN if there are no valid pixels or the denominator is zero.

%---------------------------------------------------------
\subsection{\texttt{HelicityCalculator}}\classlabel{HelicityCalculator}
Computes three moments of the current-helicity proxy $h_c = B_z J_z$ (the vertical contribution to $\mathbf{B}\cdot\nabla\times\mathbf{B}$), in G$^2$\,m$^{-1}$:
\[ \textsc{meanjzh} = \frac{1}{N}\sum_i h_{c,i}, \qquad \textsc{totusjh} = \sum_i |h_{c,i}|, \qquad \textsc{absnjzh} = \Bigl|\sum_i h_{c,i}\Bigr|, \]
with uncertainties \textsc{errmih}, \textsc{errtui}, and \textsc{errtai}. Pixels where $B_z$ or $J_z$ is effectively zero are excluded. \textsc{totusjh} measures the total amount of twist in the region irrespective of handedness. \textsc{absnjzh} measures how strongly the twist is dominated by a single handedness, which is relevant because net helicity is thought to accumulate before eruptions. As in SHARP, no Gauss-to-Tesla conversion is applied.

%---------------------------------------------------------
\subsection{\texttt{PolarityCalculator}}\classlabel{PolarityCalculator}
Computes the sum of the absolute net currents in each polarity,
\[ \textsc{savncpp} = \Bigl|\sum_{B_z>0} J_z\Bigr| + \Bigl|\sum_{B_z\le0} J_z\Bigr|, \]
converted to Amperes, with uncertainty \textsc{errjht}. In a current-neutralized region, the current flowing up through each polarity is balanced by return currents in that polarity, so each sum is close to zero. A large \textsc{savncpp} indicates non-neutralized currents, which have been linked to eruptive activity~\cite{Georgoulis:2012:non-neutralized}.

%---------------------------------------------------------
\subsection{\texttt{FreeEnergyCalculator}}\classlabel{FreeEnergyCalculator}
Computes a proxy for the photospheric excess (free) magnetic energy density from the difference between the observed and potential horizontal fields:
\begin{gather*} \rho_e = \frac{(B_x - B_x^p)^2 + (B_y - B_y^p)^2}{8\pi}, \\ \textsc{meanpot} = \frac{1}{N}\sum_i \rho_{e,i} \;\;[\mathrm{erg\,cm^{-3}}], \qquad \textsc{totpot} = \sum_i \rho_{e,i}\,dA \;\;[\mathrm{erg\,cm^{-1}}], \end{gather*}
with uncertainties \textsc{errmpot} and \textsc{errtpot}. The inputs are the observed $B_x$, $B_y$ with their errors and a potential field $B_x^p$, $B_y^p$ from \texttt{GreenpotCalculator} (to reproduce the SHARP keywords), \texttt{GreenFunctionPotentialCalculator}, or \texttt{LinearForceFreeFieldCalculator}. The choice of potential-field model therefore directly affects these values. The free energy is the energy available to be released in flares and CMEs. (The return-value descriptions in the current docstring are copied from \texttt{AlphaCalculator}; the units given here are the correct ones.)

%---------------------------------------------------------
\subsection{\texttt{ShearAngleCalculator}}\classlabel{ShearAngleCalculator}
Computes region-level statistics of the 3-D shear angle $\Theta$ between the observed and potential fields (defined in Section~\ref{cls:ShearAngleFieldCalculator}): the mean shear angle \textsc{meanshr} in degrees, its uncertainty \textsc{errmsha}, and \textsc{shrgt45}, the percentage of valid pixels with $\Theta>45^\circ$. The inputs are the observed field components and their errors and the potential-field components. Large shear concentrated along the polarity inversion line is the classic signature of a stressed, energized magnetic configuration.

The uncertainty is propagated from the errors of $B_x$, $B_y$, and $B_z$, with the potential field $B_x^p$, $B_y^p$ treated as exact. Because $B_z^p=B_z$, the observed $B_z$ enters both vectors, and its derivative includes both contributions. Writing $\mathbf{B}=(B_x,B_y,B_z)$ and $\mathbf{B}^p=(B_x^p,B_y^p,B_z)$,
\[ \frac{\partial\Theta}{\partial B_x} = -\frac{B_x^p|\mathbf{B}|^2 - (\mathbf{B}\cdot\mathbf{B}^p)\,B_x}{|\mathbf{B}|^2\,|\mathbf{B}\times\mathbf{B}^p|}, \qquad
   \frac{\partial\Theta}{\partial B_z} = -\frac{B_z\left[2|\mathbf{B}|^2|\mathbf{B}^p|^2 - (\mathbf{B}\cdot\mathbf{B}^p)(|\mathbf{B}|^2+|\mathbf{B}^p|^2)\right]}{|\mathbf{B}|^2\,|\mathbf{B}^p|^2\,|\mathbf{B}\times\mathbf{B}^p|}, \]
and $\partial\Theta/\partial B_y$ is obtained from $\partial\Theta/\partial B_x$ by exchanging $x$ and $y$. Then
\[ \textsc{errmsha} = \frac{1}{N}\Bigl[\sum_i \sum_{k\in\{x,y,z\}} \Bigl(\frac{\partial\Theta_i}{\partial B_{k,i}}\Bigr)^2 \sigma_{B_k,i}^2\Bigr]^{1/2}, \]
where $N$ is the number of pixels used for \textsc{meanshr}. $|\mathbf{B}\times\mathbf{B}^p|$ is computed from the cross product rather than as $\sqrt{|\mathbf{B}|^2|\mathbf{B}^p|^2-(\mathbf{B}\cdot\mathbf{B}^p)^2}$, to avoid cancellation where the two fields are nearly parallel. Pixels where the two fields are exactly parallel, and the derivative is singular, are left out of the sum. When no pixels are valid, \textsc{errmsha} is NaN and \textsc{meanshr} and \textsc{shrgt45} are zero.

With the \texttt{GreenpotCalculator} potential field, \textsc{meanshr} and \textsc{shrgt45} agree with the JSOC SHARP keywords to within about 0.03\% (Section~\ref{cls:GreenpotCalculator}). Those keywords therefore share the weak JSOC potential field; with a correct potential field the values differ (Section~\ref{sec:shear_weighted}). \textsc{errmsha} does not agree with the keyword, and is not meant to. The JSOC code overwrites the per-pixel error term instead of summing it, so the keyword holds the error of the last valid pixel only. It also uses the $B_x$ term in place of the $B_y$ and $B_z$ terms, and divides by the number of pixels with $\Theta>45^\circ$ rather than by $N$. For HARP~377 at 2011-02-15 02:00:00~TAI, the keyword is $6.5\times10^{-4}$\textdegree. The error propagated over all 40{,}650 valid pixels is $2.8\times10^{-2}$\textdegree, about 43 times larger.

%---------------------------------------------------------
\subsection{\texttt{LorentzCalculator}}\classlabel{LorentzCalculator}
Computes the net Lorentz force that the field above the photosphere exerts on the photospheric boundary, following Fisher et al.~\cite{Fisher2012}:
\[ F_z = \sum_i \frac{B_{x,i}^2 + B_{y,i}^2 - B_{z,i}^2}{8\pi}\,dA, \qquad F_x = -\sum_i \frac{B_{x,i}B_{z,i}}{4\pi}\,dA, \qquad F_y = -\sum_i \frac{B_{y,i}B_{z,i}}{4\pi}\,dA, \]
reported as \textsc{totfz}, \textsc{totfx} (along $\phi$), and \textsc{totfy} (along $-\theta$) in dynes. It also reports \textsc{totbsq} $=\sum_i |\mathbf{B}_i|^2$ (G$^2$) and the force components divided by \textsc{totbsq}, \textsc{epsz}, \textsc{epsx}, and \textsc{epsy}, each with a 1-$\sigma$ uncertainty. Abrupt changes in these forces are expected at the photosphere when the coronal field restructures during a flare~\cite{Fisher2012}.

%---------------------------------------------------------
\subsection{\texttt{RValueCalculator}}\classlabel{RValueCalculator}
Computes Schrijver's $R$ parameter~\cite{schrijver2007characteristic}, which measures the total unsigned flux close to strong-field polarity inversion lines. The algorithm takes the line-of-sight field and its error map and proceeds as follows: (1) the field is rebinned to approximately 2\arcsec\ resolution; (2) binary maps of strong positive ($>+150$\,G) and strong negative ($<-150$\,G) pixels are formed; (3) each map is dilated with a $3\times3$ box kernel; (4) pixels where the two dilated maps overlap mark the strong-gradient polarity inversion line; (5) this line mask is smoothed with a Gaussian of FWHM 10 pixels to form a weighting map $W$; and (6) $R=\log_{10}\sum W|B_{\mathrm{los}}|$. The output \textsc{r\_value} is dimensionless and is accompanied by \textsc{r\_value\_err}.

%---------------------------------------------------------
\subsection{\texttt{ShearWeightedRValueCalculator}}\classlabel{ShearWeightedRValueCalculator}
A variant of the $R$ parameter in which the flux near the polarity inversion line is also weighted by a per-pixel weight map, normally the shear magnitude $|S|$. The weight map is rebinned in the same way as the field, and the sum becomes $\sum W\,|B_{\mathrm{los}}|\,w$. The parameter therefore counts strong-field polarity inversion lines only to the extent that they are non-potentially sheared. The output columns are \textsc{shrw\_r\_value} and \textsc{shrw\_r\_value\_err}. The weight matrix is a required argument of this class.

%---------------------------------------------------------
\subsection{\texttt{HelicityGuessCalculator}}\classlabel{HelicityGuessCalculator}
Computes a proxy (``guess'') for the magnetic helicity budget of the region from the twist-signed shear map $S$ of \texttt{ShearAngleFieldCalculator} and the radial field. At each pixel it forms the helicity-guess density $g = S\,|B_z|$ and a polarity-weighted version $g_p = |g|\,\operatorname{sign}(B_z)$. Pixels with $|g|$ below the constructor's \texttt{threshold} are set to zero. Each quantity is then summed over the region and multiplied by $dA$ (giving G\,cm$^2$), separating the contributions by sign:
\begin{itemize}[nosep]
    \item \textsc{lhel}, \textsc{rhel}: sums of $g$ over pixels with left-handed ($S<0$) and right-handed ($S>0$) twist;
    \item \textsc{unsgnhel}, \textsc{nethel}: the unsigned ($\sum|g|$) and net ($\sum g$) totals;
    \item \textsc{pospol}, \textsc{negpol}, \textsc{unsgnpol}, \textsc{netpol}: the same decomposition for $g_p$.
\end{itemize}
Each sum has a 1-$\sigma$ uncertainty (\textsc{err}\dots), propagated from $\sigma_{B_z}$ only, with the shear treated as exact. The left- and right-handed sums describe how much flux carries each sense of twist. A region dominated by one handedness has $|\textsc{nethel}|\approx\textsc{unsgnhel}$.

%////////////////////Section\\\\\\\\\\\\\\\\\\\\\\\\\\\\\\
\section{Shear-weighted parameters and the potential-field model}\label{sec:shear_weighted}

Every calculator in this chapter that accepts a \texttt{weight\_matrix} can produce a shear-weighted version of its parameter. In the SWAN-SF-V2 dataset, the weight at each pixel is the shear magnitude
\[ w = |S| = |\sin\Theta| \quad (w=0 \text{ where } |B_z|<100\,\mathrm{G}), \]
where $S$ is the output of \texttt{ShearAngleFieldCalculator.calc} applied to $(B_x, B_y, B_z, J_z, B_x^p, B_y^p)$. The same $w$ is passed to every calculator. For example, the weighted unsigned flux becomes $\sum_i w_i|B_{z,i}|\,dA$ and the weighted total current becomes $\sum_i w_i|J_{z,i}|\,dA$. A weighted parameter counts only the non-potential part of the region: flux or current in field that is close to potential contributes little, and field sheared by $90^\circ$ contributes fully. Because $w$ is measured relative to a potential field, every weighted parameter depends on the potential-field model, including parameters such as \textsc{usflux} and \textsc{totusjz} whose unweighted definitions do not involve the potential field at all.

\begin{figure}[htb!]
    \centering
    \includegraphics[width=\textwidth]{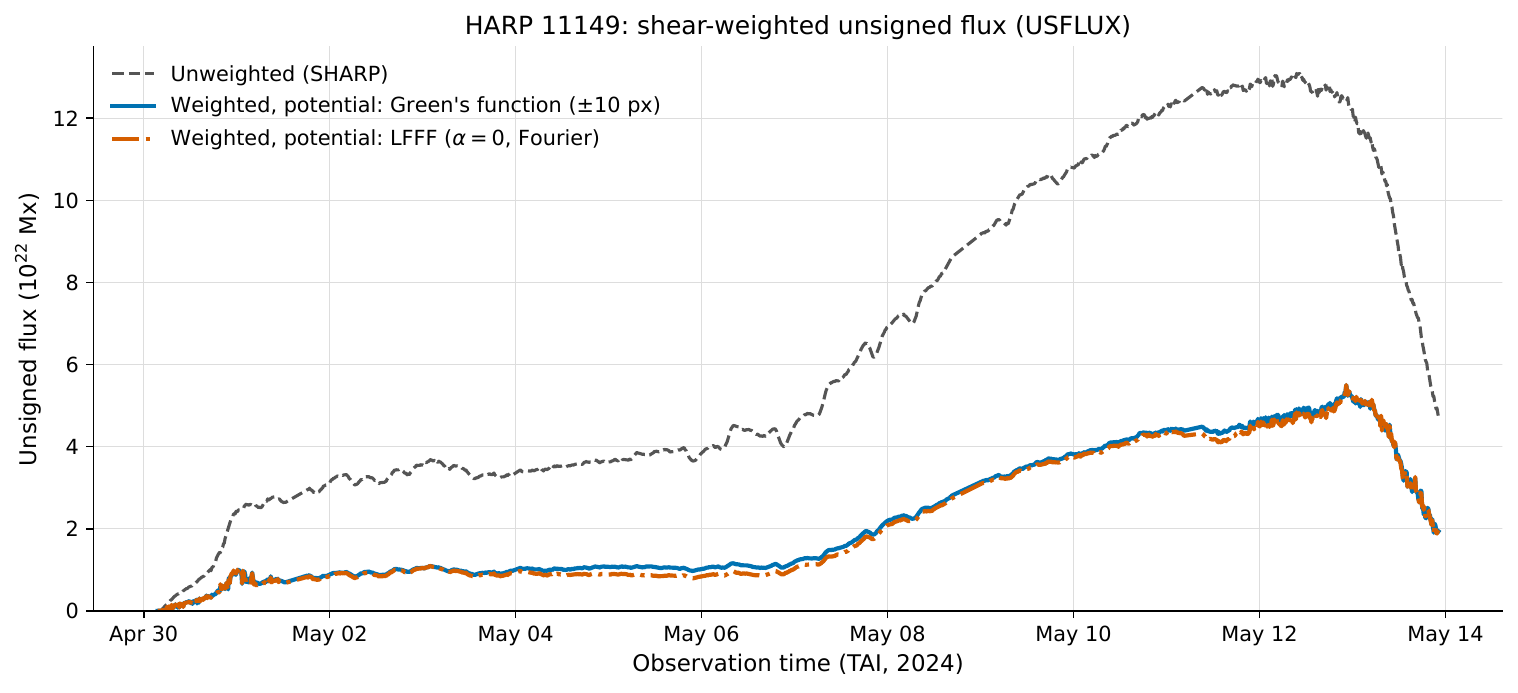}
    \caption{Unsigned flux (\textsc{usflux}, \texttt{FluxCalculator}) of HARP~11149 over its disk passage: unweighted (dashed gray), and weighted by the shear magnitude $|\sin\Theta|$ computed with the \texttt{GreenFunctionPotentialCalculator} ($\pm10$-pixel window; solid blue) and \texttt{LinearForceFreeFieldCalculator} with $\alpha=0$ (dash-dot orange) potential fields. Every 12-minute record is used.}
    \label{fig:unsflx_weighted}
\end{figure}

\begin{figure}[htb!]
    \centering
    \includegraphics[width=\textwidth]{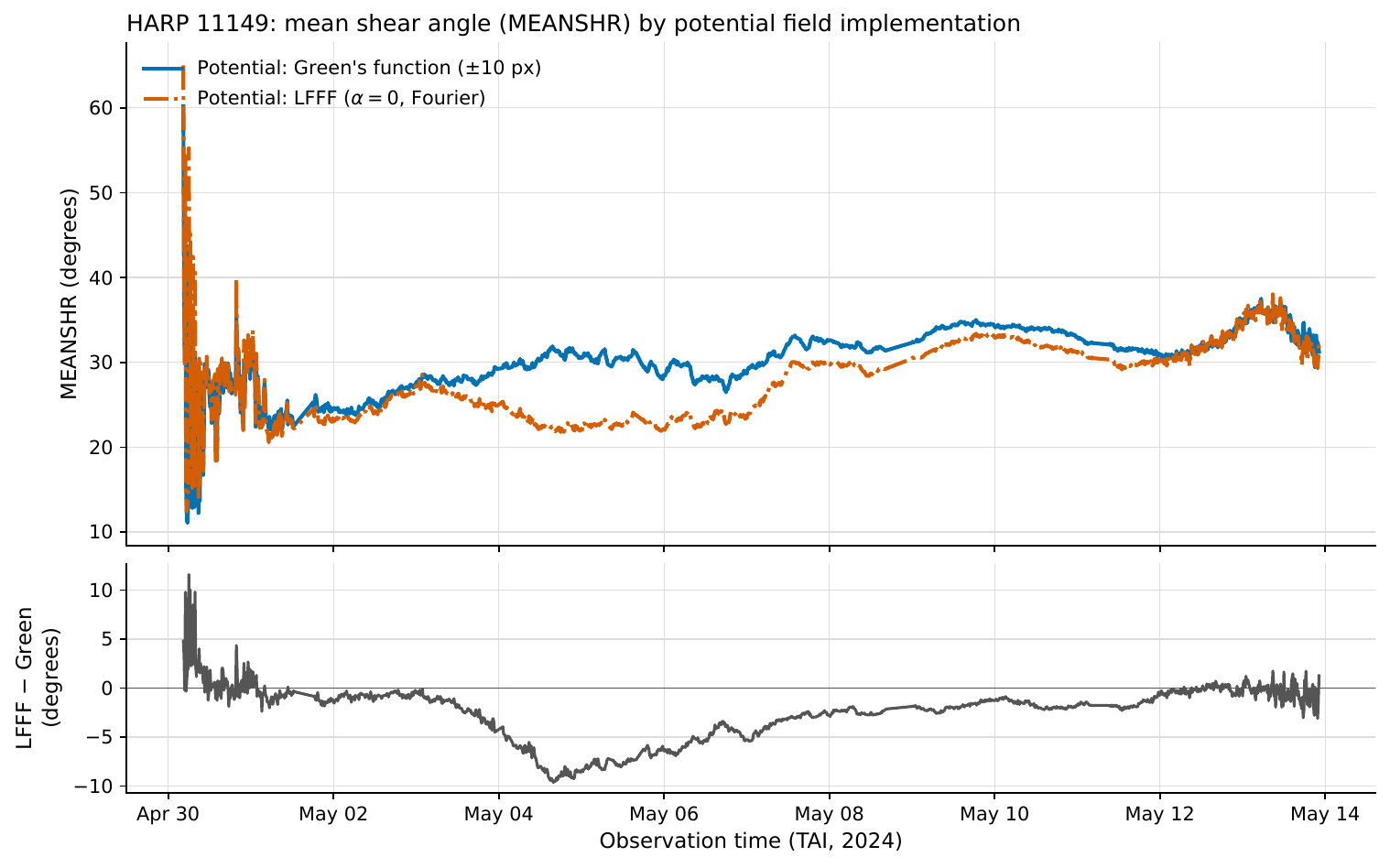}
    \caption{Mean shear angle (\textsc{meanshr}, \texttt{ShearAngleCalculator}) of HARP~11149 over its disk passage, computed with the potential field from \texttt{GreenFunctionPotentialCalculator} ($\pm10$-pixel window; solid blue) and \texttt{LinearForceFreeFieldCalculator} with $\alpha=0$ (dash-dot orange). The lower panel shows the difference, Fourier minus Green's function. No weight matrix is used. Every 12-minute record is used.}
    \label{fig:meanshr_compare}
\end{figure}

Figures~\ref{fig:unsflx_weighted} and~\ref{fig:meanshr_compare} show this dependence over the disk passage of HARP~11149 (2024-04-30 03:36 to 2024-05-13 23:00~TAI, 1538 records at the native 12-minute cadence). The first figure shows a parameter whose dependence on the potential field comes only from the weight. The second shows a parameter that is computed from the potential field directly.

\paragraph{Weighted unsigned flux (Figure~\ref{fig:unsflx_weighted}).} All three curves follow the emergence and growth of the region to a broad maximum on 12~May (unweighted peak $1.31\times10^{23}$\,Mx). The Green's-function-weighted flux is typically 30\% of the unweighted flux (interquartile range 28--35\%) and the Fourier-weighted flux typically 29\%; both peak at $5.5\times10^{22}$\,Mx. The two weighted series are almost perfectly correlated (0.999), and the median Fourier/Green's-function ratio is 0.97 (interquartile range 0.90--0.98). The ratio is lowest, 0.82--0.87, on 4--6~May, while the large sunspots form: there the $\pm10$-pixel window underestimates the horizontal potential field of the spots (Section~\ref{sec:potential_model_comparison}), so the Green's-function weight counts more of the spot flux as sheared. With the \texttt{GreenpotCalculator} potential field the dependence on the model is much larger. In an earlier run of the same comparison, the \texttt{GreenpotCalculator}-weighted flux was typically 44\% of the unweighted flux, the median Fourier/\texttt{GreenpotCalculator} ratio was 0.76 (interquartile range 0.62--0.85), and the correlation of the two weighted series was 0.975.

\paragraph{Mean shear angle (Figure~\ref{fig:meanshr_compare}).} \textsc{meanshr} (\texttt{ShearAngleCalculator}, Section~\ref{cls:ShearAngleCalculator}) is the mean angle between the observed and potential fields. Its unweighted definition already depends on the potential field, so it shows the effect of the model choice directly. The two curves agree at the start of the passage (about 23--25\textdegree\ on 1--2~May), while the region is small. They then separate while the sunspots form: on 4--6~May the Green's-function \textsc{meanshr} is about 28--31\textdegree, and the Fourier \textsc{meanshr} about 22--23\textdegree. The difference reaches $-9.6$\textdegree\ on 4~May. From 7~May the curves converge again, and both rise to 32--35\textdegree\ on 9--10~May and 13~May, while the region is producing its X-class flares. Over the passage the mean difference is $-2.5$\textdegree\ and the correlation between the two series is 0.80. The fraction of strongly sheared area (\textsc{shrgt45}) differs little: its median is 21.2\% with the Green's-function model and 19.3\% with the Fourier model. With \texttt{GreenpotCalculator} the earlier run gave a very different picture: \textsc{meanshr} rose to about 49\textdegree\ on 7--8~May, the difference from the Fourier value reached $-25$\textdegree, the correlation between the two series was 0.17, and the median \textsc{shrgt45} was 37.5\%. Much of the apparent model dependence of the SHARP shear parameters therefore comes from the JSOC potential field, not from the choice between a Green's-function and a Fourier solution. A remaining difference, concentrated while the spots form, comes from the truncated $\pm10$-pixel window.

Shear-weighted and shear-based parameters computed with different potential-field models should therefore not be mixed in one dataset, and the model used should be reported with the data. The steep decline in the flux curves on 13~May, and the scatter in \textsc{meanshr} on 30~April, coincide with the region being close to the limb, where projection effects and the quality mask remove most pixels. They are probably not real changes in the region. Records with no valid pixels are reported as zero flux and are omitted from the \textsc{meanshr} curves.

%////////////////////Section\\\\\\\\\\\\\\\\\\\\\\\\\\\\\\
\section{Effective Magnetic Field and Magnetic Charge Topology (\texttt{magparam.mct})}\label{sec:mct}

The calculators in Section~\ref{sec:magparam} reduce per-pixel field maps directly to sums and means. The \texttt{magparam.mct} sub-package instead computes a parameter from a \emph{model} of the region's magnetic connectivity: the effective connected magnetic field $B_\text{eff}$ of Georgoulis \& Rust~\cite{Georgoulis_2007}. The photospheric field is partitioned into discrete flux concentrations (``magnetic charges''), following the Magnetic Charge Topology (MCT) model of Barnes et al.~\cite{Barnes_2005}. Positive and negative concentrations are then connected so that the total length of the flux connections is approximately minimized, and $B_\text{eff}$ sums the connected flux weighted by the inverse square of the connection length:
\begin{equation}
    B_\text{eff} = \sum_{i} \frac{\Phi_i}{\left(L_i\,\lambda\right)^2},
    \label{eq:beff}
\end{equation}
where $\Phi_i$ is the flux (Mx) shared by the $i$-th connected pair of opposite-polarity partitions, $L_i$ is the distance between the two partitions' flux-weighted centroids (in pixels), and $\lambda$ is the physical size of a pixel (cm). Strong flux that closes over short distances across a polarity inversion line contributes the most. This is the configuration in which large amounts of free magnetic energy can be stored, and $B_\text{eff}$ has been shown to be a strong discriminator of flaring and non-flaring regions~\cite{Georgoulis_2007}. The sub-package is a Python port of the C++ MCT and $B_\text{eff}$ code of the \texttt{SEP-Mag-Feature-Processor} project, rewritten to operate on 2-D \texttt{numpy} arrays.

%---------------------------------------------------------
\subsection{\texttt{BeffCalculator}}\classlabel{BeffCalculator}

\texttt{BeffCalculator} is the entry point of the sub-package. \texttt{calculate(bz, cdelt1)} takes a 2-D map of the radial (or line-of-sight) field in Gauss and its plate scale in arcseconds per pixel. For SHARP CEA data, \textsc{cdelt1} must be converted from degrees, as described in Section~\ref{sec:magparam}. The calculation has three steps:
\begin{enumerate}[nosep]
    \item The map is downsampled to \texttt{scale} percent of its size by average pooling. Each output pixel is the mean of the in-bounds input pixels in a window of odd width $\approx 1+1/s$ around the corresponding source pixel, where $s=\texttt{scale}/100$. Downsampling suppresses small-scale structure and greatly reduces the run time of the connectivity step, whose cost grows quickly with the number of spread flux units.
    \item The downsampled map is passed to \texttt{SepLengthCalculator}, which builds the MCT model and the partition connections (Sections~\ref{cls:MCTPartitioner}--\ref{cls:SepLengthCalculator}).
    \item Equation~\ref{eq:beff} is evaluated over the connected pairs, with $\lambda = \texttt{len\_per\_arc\_sec}\cdot\textsc{cdelt1}/s$, the side length of a downsampled pixel in cm.
\end{enumerate}
The constructor takes the \texttt{SepLengthCalculator}, \texttt{len\_per\_arc\_sec} (the physical length of one arcsecond on the Sun, about $7.25\times10^{7}$\,cm at 1\,AU), and \texttt{scale}. The method returns a single-row \texttt{DataFrame}:
\begin{itemize}[nosep]
    \item \textsc{beff}: the effective connected magnetic field (G), Eq.~\ref{eq:beff};
    \item \textsc{beff\_max}: the upper bound obtained if every connected pair were one pixel apart, $\sum_i\Phi_i/\lambda^2$ (G);
    \item \textsc{sep\_len}: the effective separation length $L_\text{eff}\,\lambda$ (cm), defined in Section~\ref{cls:SepLengthCalculator};
    \item \textsc{mean\_flux}: the mean of the total positive and total negative flux (Mx);
    \item \textsc{err\_flux}: the flux imbalance between the two polarities (Mx).
\end{itemize}
If only one polarity is present, or no flux can be connected, \textsc{beff} and \textsc{beff\_max} are zero. The pipeline does not mask invalid pixels, so NaNs in the input should be replaced (for example by zero) before calling \texttt{calculate}.

%---------------------------------------------------------
\subsection{Building the MCT model}\classlabel{MCTModelCreator}\classlabel{MCTPartitioner}\classlabel{MCTModel}

\texttt{MCTModelCreator.process(bz, cdelt1)} turns a field map into an \texttt{MCTModel}. First, spike pixels whose absolute value exceeds \texttt{cleaning\_threshold} (G) are replaced by the mean of their $3\times3$ neighbourhood. The cleaned map is then partitioned by an \texttt{MCTPartitioner}. Finally, the flux-weighted centroid and the total uncanceled (signed) flux, in Mx, are computed for each partition. The returned \texttt{MCTModel} holds the number of partitions \texttt{n\_elem}, the \texttt{partitions} (lists of flat pixel indices), an $(n_\text{elem},2)$ array of \texttt{centroids} in (row, column) coordinates, and the signed \texttt{uncanceled\_flux\_values}.

\texttt{MCTPartitioner} runs the five-stage partitioning pipeline described below. Its constructor takes the five stage objects, a box-smoothing width \texttt{fwidth}, and a minimum partition size \texttt{min\_size} (pixels):
\begin{enumerate}[nosep]
    \item \textbf{initial clustering} of strong-field pixels of each polarity (\texttt{ClusterPartitioner});
    \item \textbf{downhill sub-partitioning}, which splits each cluster around its local flux peaks after smoothing (\texttt{DownhillClusterPartitioner});
    \item \textbf{island absorption}, which merges partitions entirely enclosed by same-polarity partitions (\texttt{IslandSwallower});
    \item \textbf{connectivity correction}, which splits any partition that is no longer one connected piece (\texttt{ClusterCorrector});
    \item \textbf{flux and size filtering}, which removes partitions with too little flux (using \texttt{SmallFlux\-Structure\-Remover}) or with fewer than \texttt{min\_size} pixels.
\end{enumerate}

%---------------------------------------------------------
\subsection{Partitioning stages}\classlabel{ClusterPartitioner}\classlabel{ClusterConnector}\classlabel{DownhillClusterPartitioner}\classlabel{IslandSwallower}\classlabel{ClusterCorrector}\classlabel{SmallFluxStructureRemover}

\begin{itemize}
    \item \textbf{\texttt{ClusterPartitioner}} smooths the map with an \texttt{fwidth}$\times$\texttt{fwidth} box filter. Near the edges only in-bounds pixels are averaged, so no zero padding is introduced. Smoothed pixels above $+$\texttt{bz\_threshold} are collected as positive candidates and those below $-$\texttt{bz\_threshold} as negative candidates. Each polarity is then split into 4-connected clusters, clusters smaller than \texttt{min\_size} pixels are dropped, and the rest are returned from largest to smallest. The threshold sets which field is considered part of a magnetic charge: raising it restricts the model to stronger field.
    \item \textbf{\texttt{ClusterConnector}} is a utility shared by the other stages. It splits a set of pixels into 4-connected components by breadth-first flood fill, in the same pixel order as the original implementation.
    \item \textbf{\texttt{DownhillClusterPartitioner}} divides a cluster that contains several flux peaks. A partition is seeded at the strongest unclaimed pixel and grown by absorbing 8-connected neighbours whose absolute field is positive and no larger than that of the pixel they are reached from, so it moves ``downhill'' from the peak. When it cannot grow further, a new partition is seeded at the next strongest unclaimed pixel, until every pixel in the cluster is claimed. Each local maximum of $|B_z|$ thus becomes a separate magnetic charge.
    \item \textbf{\texttt{IslandSwallower}} merges partitions whose entire boundary touches other partitions of the same polarity. Such an island is absorbed by peeling off its edge pixels one layer at a time and giving each to a neighbouring same-polarity partition. This removes spurious charges created by small peaks inside a larger flux concentration.
    \item \textbf{\texttt{ClusterCorrector}} checks every partition for connectivity after the earlier stages. A partition that has broken into several pieces is replaced by those pieces that individually meet the minimum flux and size requirements. The check is repeated until no further splits occur.
    \item \textbf{\texttt{SmallFluxStructureRemover}} removes partitions whose total unsigned flux $\sum|B|\,\lambda^2$ is below \texttt{min\_flux} (Mx). It is used both inside \texttt{ClusterCorrector} and as the final filter.
\end{itemize}

%---------------------------------------------------------
\subsection{Connecting the charges}\classlabel{FluxSpreader}\classlabel{GreedyOptimizer}\classlabel{SepLengthCalculator}\classlabel{SepLength}\classlabel{PathMap}\classlabel{FluxDistributions}

Finding the connectivity is treated as a flux-transport problem: each unit of positive flux is paired with a unit of negative flux so that the total distance between paired units is small.

\begin{itemize}
    \item \textbf{\texttt{FluxSpreader}} quantizes each partition's flux into whole units of \texttt{min\_flux} (Mx). It represents each unit by one pixel, claiming pixels outward in concentric square rings from the partition's centroid and skipping pixels already claimed by another partition. The result is a compact proxy region for each charge whose pixel count is proportional to its flux. It is returned as a \texttt{FluxDistributions} object, which holds the positive and negative regions as $(n,2)$ coordinate arrays and a 2-D map from each pixel to the partition it was spread from. A smaller \texttt{min\_flux} gives a finer mapping at a higher computational cost.
    \item \textbf{\texttt{GreedyOptimizer}} builds an approximately least-cost one-to-one mapping between positive and negative spread pixels. Region pairs are ordered by the distance between their minimum bounding rectangles. Pixel pairs from the closest region pairs are placed in a priority queue, and the closest pair whose two pixels are both still unmapped is repeatedly accepted. The pixel pairs of a more distant region pair are added only when the queue's next distance exceeds that region pair's bounding-rectangle distance. This is a greedy approximation to the optimal-transport solution. It is computed with vectorized distance arrays and a queue that holds only the best remaining candidate of each positive pixel. The result is a \texttt{PathMap} of mapped (positive, negative) pixel coordinates, in the order the mappings were made.
    \item \textbf{\texttt{SepLengthCalculator}} combines the stages. It builds the MCT model, spreads the flux, runs the optimizer, and attributes each pixel mapping to the pair of partitions it came from. This gives the flux $\Phi_i$ shared by every connected pair. It then computes the effective separation length
    \[ L_\text{eff} = \sqrt{\frac{\sum_i \Phi_i}{\sum_i \Phi_i/L_i^2}}, \]
    the single separation that gives the same $1/L^2$-weighted sum as the actual connections, so that $B_\text{eff}=\sum_i\Phi_i/(L_\text{eff}\lambda)^2$. If exactly two opposite-polarity partitions are found, $L_\text{eff}$ is the distance between their centroids. If there are fewer than two partitions, or no flux can be connected, an empty result with all values zero is returned. \texttt{find\_sep\_length(bz, cdelt1)} returns a \texttt{SepLength} object containing \texttt{sep\_len} (pixels), \texttt{mean\_flux} and \texttt{err\_flux} (Mx), and the connected pairs as parallel arrays \texttt{pos\_centroids}, \texttt{neg\_centroids}, and \texttt{pair\_flux}.
\end{itemize}

%---------------------------------------------------------
\subsection{Worked example}\label{sec:beff_example}

All of the classes are configured at construction and passed to one another (dependency injection), so a pipeline is assembled once and then reused for every magnetogram. The code below is taken from the example notebook \texttt{examples/beff\_mct\_visualization.ipynb}.

\begin{verbatim}
from swdatatoolkit.magparam.mct import (
    ClusterConnector, ClusterCorrector, ClusterPartitioner,
    DownhillClusterPartitioner, IslandSwallower, SmallFluxStructureRemover,
    MCTPartitioner, MCTModelCreator, FluxSpreader, GreedyOptimizer,
    SepLengthCalculator, BeffCalculator)

BZ_THRESHOLD, MIN_SIZE, FWIDTH = 50.0, 3, 3   # G, pixels, pixels
DS_SCALE        = 50.0                         # percent
MIN_FLUX        = 4e19                         # Mx per spread pixel
LEN_PER_ARC_SEC = 7.25e7                       # cm per arcsec

connector     = ClusterConnector()
cluster_part  = ClusterPartitioner(connector, FWIDTH, BZ_THRESHOLD, MIN_SIZE)
remover       = SmallFluxStructureRemover(0.0, LEN_PER_ARC_SEC)
corrector     = ClusterCorrector(connector, remover, MIN_SIZE)
partitioner   = MCTPartitioner(cluster_part, DownhillClusterPartitioner(),
                               IslandSwallower(), remover, corrector,
                               FWIDTH, MIN_SIZE)
model_creator = MCTModelCreator(partitioner, 4000.0, LEN_PER_ARC_SEC)
sep_calc      = SepLengthCalculator(model_creator, GreedyOptimizer(),
                                    FluxSpreader(MIN_FLUX), MIN_FLUX)
beff_calc     = BeffCalculator(sep_calc, LEN_PER_ARC_SEC, DS_SCALE)

result = beff_calc.calculate(bz, cdelt1)       # bz: 2-D array in G
print(result[["BEFF", "BEFF_MAX", "SEP_LEN"]])
\end{verbatim}

\begin{figure}[htbp]
    \centering
    \includegraphics[width=\textwidth]{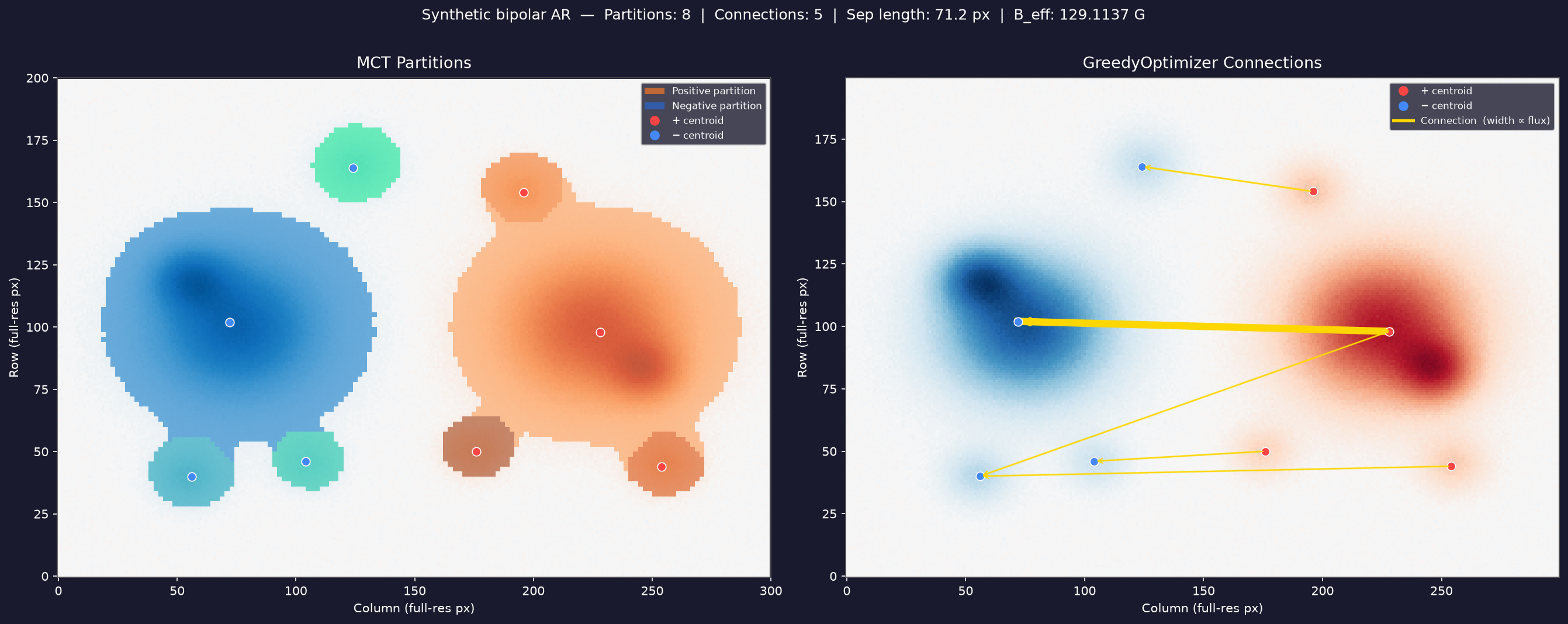}
    \caption{MCT partitions and flux connections computed by the \texttt{magparam.mct} pipeline for a synthetic bipolar active region (\texttt{examples/beff\_mct\_visualization.ipynb}). \textbf{Left:} the input magnetogram with the eight MCT partitions found on the 50\%-downsampled map (warm colors: positive; cool colors: negative) and their flux-weighted centroids. \textbf{Right:} the five partition connections found by \texttt{GreedyOptimizer}; line width is proportional to the connected flux. $B_\text{eff}=129.1$\,G and $L_\text{eff}=71.2$ downsampled pixels.}
    \label{fig:mct_viz}
\end{figure}

The notebook applies this pipeline to a synthetic $200\times300$-pixel bipolar active region (0.5\arcsec\ pixels), built from two dominant opposite-polarity Gaussian flux concentrations, six smaller satellite spots, and Gaussian noise with a standard deviation of 8\,G. Figure~\ref{fig:mct_viz} shows the result. After downsampling to 50\%, the model has eight partitions (four of each polarity) and five connected pairs. Most of the connected flux, $3.1\times10^{21}$\,Mx, runs between the two main concentrations, about 78 downsampled pixels apart. The other four connections, which link satellite spots to each other and to the main concentrations, carry $4$--$12\times10^{19}$\,Mx each. The resulting effective separation is 71.2 downsampled pixels ($5.16\times10^{9}$\,cm), and $B_\text{eff}=129.1$\,G. Because $B_\text{eff}$ weights each connection by $1/L^2$, the two short satellite connections (about 36 pixels) contribute more per unit flux than the dominant long-range connection.

%==========================================================
\chapter{Feature Extraction and Scalable Analytics}\label{chap:features}
%==========================================================

This chapter describes the classes that turn the per-timestep parameters of the previous chapters into analysis-ready samples. The \texttt{slicer} module cuts each active region's multivariate time series into labelled observation windows (Section~\ref{sec:slicer}). The \texttt{features} module then provides two classes for extracting statistical summary features from each window. Both are designed for distributed processing with Apache PySpark on the large datasets used on HPC clusters, and \texttt{StatisticalFeatures} can also be used on its own in single-node mode, for development and small datasets.

%////////////////////Section\\\\\\\\\\\\\\\\\\\\\\\\\\\\\\
\section{\texttt{HarpDataSlicer} in \texttt{slicer}}\classlabel{HarpDataSlicer}\label{sec:slicer}

Flare-forecasting models are trained on \emph{samples}, not on continuous time series. A sample is a fixed-length window of observations of one active region, paired with a label describing the flaring activity that follows it. The \texttt{slicer} module contains one class, \texttt{HarpDataSlicer}, which cuts the per-HARP multivariate time series (e.g.\ SHARP and AIA parameters at 12-minute cadence, merged with a flare catalog) into such samples using PySpark. This is the sampling scheme of the SWAN-SF benchmark dataset~\cite{angryk2020multivariate}.

\paragraph{Windows.} Each time series, one per value of \texttt{group\_col} (default \texttt{harp\_number}), is ordered by \texttt{time\_col} (default \texttt{obs\_start}) and indexed by row. A slice starts at every \texttt{stride\_steps}-th row $r$ and consists of three consecutive parts (Figure~\ref{fig:slicer}):
\begin{itemize}[nosep]
    \item an \emph{observation window} of $o=$~\texttt{obs\_steps} rows, $[r,\ r+o-1]$, which holds the input features of the sample;
    \item a \emph{latency gap} of $\tau=$~\texttt{latency\_steps} rows, which models the delay between the last observation and the issue time of a forecast;
    \item a \emph{forecast window} of up to $p=$~\texttt{pred\_steps} rows, $[r+o+\tau,\ r+o+\tau+p-1]$, which is used only to assign the label.
\end{itemize}
The defaults, $o=60$, $p=120$, and a stride of 5 rows, correspond at 12-minute cadence to a 12-hour observation window, a 24-hour forecast window, and a new sample every hour. Near the end of a series the forecast window is clipped to the last available row. A slice is kept only if the whole observation window exists and at least $\max(1,\lceil p\cdot\texttt{min\_pred\_fraction}\rceil)$ forecast rows exist. Windows are counted in rows rather than in time, so the input should be on a regular cadence: gaps are not detected and stretch any window that spans them.

\paragraph{Labels.} The flare events are read from the rows of the input that have a non-null \texttt{flare\_time\_col} (default \texttt{peak\_time}) and \texttt{flare\_class\_col} (default \texttt{goes\_class}). A flare only has to be joined to some row of its HARP. The label of a slice, \texttt{peak\_class}, is the strongest flare whose peak falls inside the forecast window. Flares are ranked by GOES class letter (X~$>$~M~$>$~C~$>$~B~$>$~A) and then by magnitude, so X1.4 outranks M6.9. Slices with no flare in the forecast window are labelled \texttt{FQ} (flare quiet). The full GOES class is kept, so the caller can choose the binary or multi-class target, for example $\geq$M1.0 against everything else.

\paragraph{Input and output.} \texttt{make\_slice\_rows(merged\_df)} takes a PySpark \texttt{DataFrame} with one row per observation per HARP, containing the group, time, flare-time, and flare-class columns plus any number of parameter columns, which are carried through unchanged. It returns the data in \emph{long} format: every observation row is repeated once for each slice whose observation window contains it, and rows that belong to no valid slice are dropped. The added columns are \texttt{slice\_id} (\texttt{"<harp>\_<slice start time>"}, unique per sample), the window boundaries \texttt{slice\_start\_time}, \texttt{obs\_end\_time}, \texttt{pred\_start\_time}, and \texttt{pred\_end\_time}, and the label \texttt{peak\_class}. Grouping the output by \texttt{slice\_id} recovers the individual samples. That grouped frame is exactly the input expected by \texttt{PySparkFeatureExtractor} (Section~\ref{cls:PySparkFeatureExtractor}), so slicing and feature extraction form a single distributed pipeline (see the example in Section~\ref{cls:PySparkFeatureExtractor}).
The constructor raises \texttt{ValueError} if \texttt{obs\_steps}, \texttt{pred\_steps}, or \texttt{stride\_steps} is not positive, if \texttt{latency\_steps} is negative, or if \texttt{min\_pred\_fraction} is outside $[0,1]$.

\begin{figure}[htb!]
\centering
\begin{tikzpicture}[x=0.62cm,y=0.62cm, every node/.style={font=\footnotesize}]
  % row ticks
  \foreach \i in {0,...,18} { \draw[gray!60] (\i,0) rectangle (\i+1,0.7); }
  \node[anchor=east] at (-0.2,0.35) {rows};
  % slice 1
  \fill[roi!35] (2,1.2) rectangle (7,1.9);   \node at (4.5,1.55) {obs ($o$)};
  \fill[gray!25] (7,1.2) rectangle (9,1.9);  \node at (8,1.55) {$\tau$};
  \fill[ext!35] (9,1.2) rectangle (16,1.9);  \node at (12.5,1.55) {forecast ($p$)};
  \node[anchor=east] at (1.8,1.55) {slice at $r$};
  % slice 2
  \fill[roi!35] (4,2.3) rectangle (9,3.0);   \node at (6.5,2.65) {obs};
  \fill[gray!25] (9,2.3) rectangle (11,3.0); \node at (10,2.65) {$\tau$};
  \fill[ext!35] (11,2.3) rectangle (18,3.0); \node at (14.5,2.65) {forecast};
  \node[anchor=east] at (3.8,2.65) {slice at $r+s$};
  \draw[<->] (2,3.4) -- (4,3.4) node[midway,above] {stride $s$};
  \draw[->,ext] (13,0.8) -- (13,1.15);
  \node[ext,anchor=north] at (13,0) {flare peak $\Rightarrow$ label of slice at $r$};
\end{tikzpicture}
\caption{Windows used by \texttt{HarpDataSlicer}. Each slice consists of an observation window of $o$ rows, a latency gap of $\tau$ rows, and a forecast window of up to $p$ rows. Slices start every $s$ rows, so consecutive observation windows overlap when $s<o$. A slice is labelled with the strongest flare peaking inside its forecast window, or \texttt{FQ} if there is none.}
\label{fig:slicer}
\end{figure}

%////////////////////Section\\\\\\\\\\\\\\\\\\\\\\\\\\\\\\
\section{\texttt{StatisticalFeatures}}\classlabel{StatisticalFeatures}

Each sample produced by the slicer is a short multivariate time series. Most classical machine-learning models need a fixed-length feature vector instead, so each parameter's time series within a window is summarized by a set of statistics. \texttt{StatisticalFeatures} defines the names of the ten available summary features, and a matching set of per-series functions (\texttt{calculate\_mean()}, \texttt{calculate\_slope()}, \dots) implements them on a \texttt{pandas.Series}:
\begin{itemize}[nosep]
    \item \texttt{MEAN}, \texttt{STD}, \texttt{MIN}, \texttt{MAX}, \texttt{VARIANCE}, and \texttt{KURTOSIS}: the level, spread, extremes, and tailedness of the parameter over the window;
    \item \texttt{SLOPE}: the least-squares linear trend against the row index, which shows whether the parameter is rising or falling (e.g.\ flux emergence);
    \item \texttt{ENTROPY}: the Shannon entropy of a fixed-range histogram of the values, which requires \texttt{n\_bins}, \texttt{min\_val}, and \texttt{max\_val};
    \item \texttt{DOMINANT\_FREQUENCY}: the frequency of the strongest Fourier component, with an optional \texttt{sampling\_rate};
    \item \texttt{SPECTRAL\_ENTROPY}: the entropy of the normalized power spectrum, which is low for nearly periodic series and high for noise-like ones, with an optional \texttt{n\_bins}.
\end{itemize}
Each function drops NaN values before computing, and returns \texttt{None} rather than raising an exception for empty or all-NaN input, or for too few points (e.g.\ a slope needs at least two). This makes them safe to use in batch pipelines where some series are incomplete. The class also provides helpers to list the available features (\texttt{get\_all\_features()}), to identify which ones take parameters (\texttt{get\_parameterized\_features()}), and to validate a requested list (\texttt{validate\_features()}). \texttt{get\_order\_dependent\_features()} lists the features that treat consecutive rows as consecutive time steps (\texttt{SLOPE}, \texttt{DOMINANT\_FREQUENCY}, and \texttt{SPECTRAL\_ENTROPY}), whose values are only meaningful if the series is in time order. The function \texttt{get\_feature\_calculator(name, feature\_config)} returns the callable for a feature name with its configuration applied.

%////////////////////Section\\\\\\\\\\\\\\\\\\\\\\\\\\\\\\
\section{\texttt{PySparkFeatureExtractor}}\classlabel{PySparkFeatureExtractor}

\texttt{PySparkFeatureExtractor} computes the statistical features for every sample of a grouped PySpark DataFrame in parallel, using \texttt{applyInPandas}, so feature extraction scales to the large number of samples in a multi-year dataset.
\begin{itemize}
    \item \textit{Configuration (constructor):} \texttt{statistical\_features}, a list of feature names from \texttt{StatisticalFeatures}, and \texttt{feature\_config}, a dictionary of per-feature settings, which is required for \texttt{entropy}. Unsupported names, or \texttt{entropy} without a configuration, raise \texttt{ValueError}.
    \item \textit{Input (\texttt{do\_extraction(grouped\_df, feature\_columns, order\_column)}):} a \texttt{GroupedData} object, for example the slicer output grouped by \texttt{slice\_id}, in which each group is one sample, and the list of numeric parameter columns to summarize. Columns that are not present are skipped with a warning. \texttt{order\_column} names the time column (e.g.\ \texttt{obs\_start} for slicer output) by which the rows of each group are sorted before the features are computed. Spark does not guarantee row order within a group, especially after the join performed by the slicer, so this argument should be given whenever an order-dependent feature is requested; if it is omitted in that case, a warning is logged.
    \item \textit{Output:} a long-format DataFrame with one row per (sample, parameter, feature) and columns \texttt{sample\_id} (the group key), \texttt{column\_name}, \texttt{feature\_name}, and \texttt{feature\_value}. If a feature cannot be computed for a sample, its value is null. Pivoting on \texttt{column\_name} and \texttt{feature\_name} gives one feature vector per sample.
\end{itemize}

\paragraph{Example usage.}

\begin{verbatim}
from pyspark.sql import SparkSession
from swdatatoolkit.slicer import HarpDataSlicer
from swdatatoolkit.features import PySparkFeatureExtractor
from swdatatoolkit.features import StatisticalFeatures as SF

spark     = SparkSession.builder.appName("SWFeatures").getOrCreate()
merged_df = spark.read.parquet("sharp_with_flares.parquet")

rows = HarpDataSlicer(obs_steps=60, pred_steps=120,
                      stride_steps=5).make_slice_rows(merged_df)

extractor = PySparkFeatureExtractor(
    statistical_features=[SF.MEAN, SF.STD, SF.SLOPE, SF.MAX])
feature_df = extractor.do_extraction(rows.groupBy("slice_id"),
                                     ["TOTUSJZ", "R_VALUE", "TOTUSJH"],
                                     order_column="obs_start")
labels = rows.select("slice_id", "peak_class").distinct()
feature_df.show(5)
\end{verbatim}

%==========================================================
\chapter{Software Design and Implementation Considerations}
%==========================================================

Although the library is primarily scientific in purpose, its internal organization reflects several sound software-engineering principles.

%////////////////////Section\\\\\\\\\\\\\\\\\\\\\\\\\\\\\\
\section{Modularity}

The separation into packages for data sources, image processing, edge detection, image parameters, magnetic-field analysis, and magnetic-parameter derivation improves maintainability and conceptual clarity.

%////////////////////Section\\\\\\\\\\\\\\\\\\\\\\\\\\\\\\
\section{Reusability}

The calculators and downloaders are structured as reusable components, allowing them to be incorporated into custom workflows without significant rewriting.

%////////////////////Section\\\\\\\\\\\\\\\\\\\\\\\\\\\\\\
\section{Extensibility}

The modular design allows new data sources, feature definitions, or magnetic calculators to be added with minimal disruption to the existing structure.

%////////////////////Section\\\\\\\\\\\\\\\\\\\\\\\\\\\\\\
\section{Reproducibility}

By consolidating common analysis routines in a shared library, \href{https://dmlab.cs.gsu.edu/docs/spaceweather_toolkit/}{\textbf{swdatatoolkit}} supports reproducible scientific workflows and reduces ambiguity in feature computation.

%==========================================================
\chapter{Scientific and Practical Impact}
%==========================================================

The primary contribution of \href{https://dmlab.cs.gsu.edu/docs/spaceweather_toolkit/}{\textbf{swdatatoolkit}} is its integration of heterogeneous space weather analysis functions into a
single library. This has several practical consequences:

\begin{itemize}
    \item It lowers the barrier to entry for solar data analysis.
    \item It reduces duplication of standard preprocessing and feature computation steps.
    \item It supports consistent derivation of parameters across studies.
    \item It facilitates machine-learning applications in space weather forecasting.
    \item It improves efficiency in building analysis-ready datasets.
\end{itemize}

\noindent For researchers working with solar imagery and magnetic data, these properties are especially valuable because they align well with the workflow demands of modern heliophysics.

%==========================================================
\chapter{Limitations and Scope of Use}
%==========================================================

As with any scientific toolkit, the practical utility of \href{https://dmlab.cs.gsu.edu/docs/spaceweather_toolkit/}{\textbf{swdatatoolkit}} depends on the suitability of its components for the research question at hand. Some functions may require domain knowledge to interpret correctly, particularly those involving magnetic extrapolation or derived physical quantities. Additionally, the validity of downstream inferences depends on the quality and consistency of the input data.\\

\noindent The toolkit should therefore be understood as an enabling framework, rather than a substitute for scientific interpretation.

%==========================================================
\chapter{Documentation Infrastructure}
%==========================================================

A fully automated documentation website for \href{https://dmlab.cs.gsu.edu/docs/spaceweather_toolkit/}{\textbf{swdatatoolkit}} was built using \href{https://www.sphinx-doc.org/}{Sphinx} and deployed via a CI/CD pipeline. This infrastructure ensures that documentation remains synchronized with the codebase and is always publicly accessible.

%////////////////////Section\\\\\\\\\\\\\\\\\\\\\\\\\\\\\\
\section{Sphinx Documentation System}

The documentation system was established from the ground up, covering:

\begin{itemize}
    \item A complete \texttt{conf.py} configuration enabling auto-API generation, OpenGraph metadata, inter-Sphinx cross-references, copy-button code blocks, and a
          custom \texttt{sunpy-sphinx-theme}-based layout.
          
    \item A structured \texttt{index.rst} entry point with module hierarchy, installation guide, tutorial pages, and API reference.
    
    \item Custom HTML templates (\texttt{layout.html}, \texttt{custom\_footer.html}) for branding and navigation.
    
    \item Tutorial pages with grid-card layouts demonstrating end-to-end usage of the library.
    
    \item A \texttt{CONTRIBUTORS.md} page listing all project contributors.
    
    \item A \texttt{TEST\_BUILD.md} guide documenting the local build and test process for developers.
    
\end{itemize}

%////////////////////Section\\\\\\\\\\\\\\\\\\\\\\\\\\\\\\
\section{CI/CD Deployment Pipeline}

A Bitbucket Pipelines configuration (\texttt{bitbucket-pipelines.yml}) was written to automatically build and deploy the Sphinx documentation on every push to the main branch. This ensures the live documentation site at \href{https://dmlab.cs.gsu.edu/docs/spaceweather_toolkit/}{\texttt{dmlab.cs.gsu.edu/docs/spaceweather\_toolkit/}} always reflects the current state of the library without any manual deployment steps.

%////////////////////Section\\\\\\\\\\\\\\\\\\\\\\\\\\\\\\
\section{Documentation Scope}

The website documents every module in the library at the API level, including full docstrings, parameter tables, and return types for all calculators, downloaders, and feature extractors described in this paper. The tutorials section demonstrates class-level usage examples.

%////////////////////Section\\\\\\\\\\\\\\\\\\\\\\\\\\\\\\
\section{Package Index Deployment}

In addition to documentation deployment, the project includes an automated package publication pipeline for Python package index distribution. This workflow provides a reproducible mechanism for validating release packaging and distribution before publication to the main Python Package Index (PyPI). By automating packaging and upload steps, the pipeline reduces manual release overhead and helps ensure that published distributions are built from a clean environment.

%==========================================================
\chapter{Conclusion}
%==========================================================

\href{https://dmlab.cs.gsu.edu/docs/spaceweather_toolkit/}{\textbf{swdatatoolkit}} is a comprehensive Python library for space weather research that integrates data acquisition, image processing, magnetic-field analysis, and feature engineering. Its available functionality supports the full pipeline from raw heliophysics observations to analysis-ready descriptors. The library is especially well suited to active-region studies, flare-related investigations, and machine-learning applications that rely on structured observational features.
\\

\noindent In summary, \href{https://dmlab.cs.gsu.edu/docs/spaceweather_toolkit/}{\textbf{swdatatoolkit}} provides a practical and scientifically grounded foundation for reproducible solar data analysis.

%==========================================================
\chapter*{Acknowledgment}
%==========================================================

\href{https://dmlab.cs.gsu.edu/docs/spaceweather_toolkit/}{\textbf{swdatatoolkit}} is associated with the \href{https://dmlab.cs.gsu.edu}{Data Mining Lab} at \href{https://gsu.edu}{Georgia State University}, and is positioned within a broader research context involving solar physics and space weather data science.
\bigskip

\noindent This project has been supported in part by funding from the Division of Advanced Cyberinfrastructure within the Directorate for Computer and Information Science and Engineering, the Division of Atmospheric \& Geospace Sciences within the Directorate for Geosciences, under NSF awards \#1931555 and \#1936361. It has also been supported by NASA's Space Weather Science Application Research-to-Operations-to-Research program grant \#80NSSC22K0272, Multidomain Reusable Artificial Intelligence Tools program grant \#80NSSC23K1026, and Heliophysics Artificial Intelligence/Machine Learning-ready Data program grant \#80NSSC24K0238.

\printbibliography

\end{document}